\documentclass[%
reprint,
 amsmath,amssymb,
 aps,
prd,
floatfix,
showpacs,
]{revtex4-2}

\usepackage{comment}
\usepackage{graphicx}
\usepackage{dcolumn}
\usepackage{bm}

\begin{document}


\preprint{}

\title{Magnetohydrodynamic waves excited by a coupling between gravitational waves and a strongly magnetized plasma in binaries of neutron stars}

\author{Adam S. Gontijo}
 \altaffiliation{adam.assis@inpe.br}
\author{Oswaldo D. Miranda}%
 \email{oswaldo.miranda@inpe.br}
\affiliation{%
 INPE --- Instituto Nacional de Pesquisas Espaciais --- Divis\~ao de Astrof\'isica, \\
 S\~ao Jos\'e dos Campos, 12227-010 S\~ao Paulo, Brazil
}%

\date{\today}

\begin{abstract}
Coalescence of binary neutron stars (BNSs) is one of the sources of gravitational waves (GWs) able to be detected by ground-based interferometric detectors. The event GW170817 was the first observed in the gravitational and electromagnetic spectra, showing through this joint analysis a certain compatibility with the models of short gamma-ray bursts (sGRBs) to explain the signature of this system. Due to the intense magnetic fields of the neutron stars, the plasma magnetosphere stays strongly magnetized and the propagation of the GW through plasma can excite magnetohydrodynamic (MHD) modes such as Alfv\'en and magnetosonic waves. The MHD modes carry energy and momentum through the plasma, suggesting a mechanism to accelerate the matter during the coalescence of the binaries, explaining some characteristics of the fireball model of the sGRBs. We present a semianalytical formalism to determine the energy transferred by the GW-MHD interaction during the inspiral phase of the stars. Using the inferred physical parameters for GW170817 and considering that the magnetic fields on the surfaces of the stars are $10^{8}\, {\rm T}$, we show that the energy in the plasma can reach maximum value $\sim 10^{35}\,{\rm J}$ ($\sim 10^{32}\,{\rm J}$) for the Alfv\'en mode (magnetosonic mode) if the angle formed between the background magnetic field and the GW propagation direction is $\theta = \pi / 4$. Particularly, for $\theta = \pi / 2$ only the magnetosonic mode is in coherence with the GWs. In this case, the excited energy in the plasma reaches maximum value $\sim 10^{36} {\rm J}$. If the magnetic field on the surface of the progenitors of the event GW170817 was $\sim 2\times 10^{9}\,{\rm T}$ then energies comparable to those inferred for the  GRB 170817A could be obtained. In particular, our semianalytical formalism show consistence with the results obtained by other authors through full general relativistic magnetohydrodynamics (GRMHD) simulations.
\end{abstract}

\pacs{04.30.−w, 97.60.Jd, 94.30.cq}
\maketitle


\section{\label{secIntroduction}Introduction}

After the inauguration of a new observational window for the study of the Universe through the gravitational waves (GWs) produced by black hole (BH) mergers \cite{1,2,3,4,5}, a sign in GWs of the event GW170817, detected by the LIGO-VIRGO collaboration, was interpreted as being consistent with the coalescence of a binary of neutron stars (BNSs) \cite{6}. This detection comes four decades after the seminal results of \cite{7} who evaluated the orbital decaying of the PSR B1913+16.

The best combination of mass measurement provides the chirp mass $M_{c}=1.188^{+0.004}_{-0.002}\,M_{\odot}$ and a total mass range $\sim 2.73 - 3.29\, M_{\odot}$ for GW170817. The masses of the components are between $0.86 - 2.26\, M_{\odot}$, according to the masses of known NSs \cite{6}.

In addition, a gamma-ray burst, GRB 170817A, was observed $1.7\,{\rm s}$ after the coalescing time of GW170817 \cite{8,9}. The combination of the LIGO-VIRGO data allowed for a precise positioning within an area of 28 deg$\,^2$ and to a distance $\sim 40^{+8}_{-14}\,{\rm Mpc}$ \cite{10}. On the other hand, the GRB 170817A was found in the galaxy ${\rm NGC} \,4993$, consistent with the location and distance given by the GW data \cite{11}.

Traditionally, the GRB progenitors are based on the time domain duration \cite{12}. Some long GRBs ($t\geq 2$s) are associated with Type Ic supernova \cite{13,14}. In its turn, observations by Swift satellite \cite{15} have revealed that some short GRBs ($t\leq 2$s) are associated with elliptical and small star-forming galaxies \cite{16}. Additionally, their progenitors have been associated with mergers of compact objects \cite{12,17,18,19,20}. Similar results were obtained by observations from Fermi \cite{21,22,23} and Integral satellites \cite{24}.

A part of the energy released by the GRB progenitor is in the form of highly relativistic jets. These structures can reach radius $\sim 10^{11}\,{\rm m}$ producing $\gamma$-rays which are characterized, nominally, as prompt emission. At a greater distance, for a range of $\sim 10^{12}\,{\rm m}$, the afterglow, another electromagnetic counterpart, is produced \cite{25,26}. Despite the extensive study, the nature of the GRB still remains as an open question \cite{18}.

One possibility is that GRBs are produced according to the so-called ``fireball model'' \cite{18,27,28,29,30}, introduced for the first time by \cite{31}. Fireballs are essentially dynamical objects, whose properties grow up quickly with time. The GRB itself is produced by the internal dissipation within the outflow, while the afterglow is produced by the external shock with the surrounding medium. These models are based on the assumption that ultrarelativistic outflows are commonly shock generators. Adding some baryon contamination in the fireball, even a small amount of baryons ($10^{-7}-10^{-5}\, {M}_{\odot}$), dragged by the radiative sphere would eventually carry the energy of the fireball, converting the initial radiative energy into kinetic energy \cite{18}.

The internal shock is in interaction with itself. These shocks can explain the rapidly varying light curves of the prompt emission. They occur within a structure composed by relativistic jets, when the outflow of the central engine takes successive shells with different Lorentz factors. Multiple shocks appear when the faster shells outgrow the slowest shells \cite{32,33,34,35}, and the fireball shock interacts with the external environment around the source. This shock explains the afterglow radiation of several wavelengths \cite{36,37,38,39,40}. The models of internal and external shocks do not depend on the nature of the central source. It is irrelevant whether the initial energy source is due to the coalescence of compact objects or to collapses produced by explosions of super massive stars \cite{41}.

The critical issue is related to the problem of particle acceleration in relativistic shocks. The usual model is the so-called ``diffusive shock acceleration model''. In this model, the particles are accelerated when they repeatedly cross a shock \cite{42}, and magnetic irregularities confine the particles near the shock \cite{43,44,45,46}. In the case of GRBs, the shocks are relativistic and particle acceleration becomes a more complicated process. Diffuse shock acceleration can not be applied because the propagation of the accelerated particles, close to the shock, can not be described as spatial diffusion \cite{47,48,49,50,51,52,53}.

Magnetic fields are the natural way of transmitting the energy of the central object with a small contribution of baryons. In an ideal magnetohydrodynamic (MHD) plasma, this can be considered as magnetic energy carrying by the outflow of the central object \cite{54}. This model is dominated by the Poynting flux, without internal shocks.

Despite instabilities, the magnetic energy in the flux can be converted into kinetic energy from plasma, then into heat or radiation. Depending on the velocity of this energy dissipation, the energy conversion may occur near the central object or outside the photosphere, if the dissipation is slow. At the same time, the dissipation of the magnetic energy reduces the total pressure, and the pressure gradient accelerates the flow out. As a result, a central magnetic mechanism can provide both acceleration and dissipation outside the photosphere as required for an efficient prompt radiation \cite{54}.

There are questions for which the internal fireball shock model does not provide satisfactory answers \cite{54,55,56,57,58}. An alternative model called electromagnetic model (EMM) was proposed. In this model, the energy of a dominant Poynting flux is dissipated directly into particles through plasma instabilities.  Although the Poynting flux is generally most directly observable, this electromagnetic energy can be transferred to kinetic energy of the plasma, which will radiate through different processes \cite{59}. Scenarios with dominant Poynting flux require a strong magnetic field ($>10^{15}\,{\rm G}$) and large rotation rate ($\Omega \sim 10^{4}\, {\rm s}^{-1}$) \cite{60}. These characteristics can be found when a NS merger forms an accretion torus around a BH and in newborn magnetars \cite{59,61,62,63,64}.

The presence of charged matter and high magnetic fields suggests the participation of Alfv\'en and magnetosonic waves \cite{65} as an important mechanism for particle acceleration, as well as processes involving magnetic reconnection \cite{66}. Given the high temperatures reached by the matter in the fireball \cite{67}, associated with intense magnetic fields, the magnetohydrodynamics (MHD) waves are produced during the coalescence of the compact binary systems, while they release gravitational waves \cite{68}.

The MHD waves can transport energy and momentum through the plasma. They depend on the characteristics of the system, such as magnetic field, local density, and also of the gravitational radiation. This would suggest that MHD waves may be a possible mechanism for accelerating matter to high Lorentz factors.

In a vacuum (flat) space-time, GWs and electromagnetic waves (EMWs) do not interact. However, the GWs and EMWs can couple on a curved background. In particular, \cite{69} finds low values for the transfer of electromagnetic energy to gravitational energy and being proportional to $G/(\pi c^3)B^2RT\sim 10^{-17}$. The authors obtain that result for a light-ray travel time $T\sim 10^{7}$ years and considering a constant interstellar magnetic field $\sim 10^{-5}\,{\rm G}$ over a scale $\sim 10$ light years.

The EMW traveling through a static electromagnetic field can be excited by GWs with the same frequency and wave vector. The efficiency of the process depends upon the square of the field as well as on the square of its linear dimension \cite{70}. In a space with a stationary magnetic field, \cite{71} verified that $hB_{0}$ is the source of the electromagnetic radiation, where $h$ is the GW amplitude that modifies the metric and, as a result, it produces stretch and compression of the lines of $B_{0}$.

The GW-EMW interaction has also been studied in \cite{72}. In this case, the interaction results from the decay of a graviton with energy $\hbar\omega_{0}$ into a photon with energy $\hbar\omega_{1}$ plus a plasmon with energy $\hbar\omega_{2}$, such that the growth rate of the plasma waves is given by $\left({k_{2}\omega_{0}}/{k_{2}\omega_{1}}\right)^2{\omega_{1}\omega_{2}}/{4}|h_{\times}|^2$ (considering $h_{+}=0$).

In \cite{73}, the authors derived the dispersion relations governing the coupling between GWs and EMWs and that propagate in the parallel and perpendicular directions to the background magnetic field. Waves propagating at an arbitrary angle with the magnetic field were not considered. The authors found that the GW-EMW coupling is related to the fact that anisotropic perturbations of the distribution function induce drifts in the electrons and ions, which in turn generate an electric current.

Therefore, the GW-EMW interaction is mainly related with the generation of electric currents in the plasma due to the perturbations in the trajectories of the charged particles by the passage of the GW through the medium \cite{74}. The GW propagation along the direction of the background magnetic field does not generate currents in the plasma \cite{73,74}.    

Although the interaction of the GWs with the matter be weak, the GW-EMW coupling can be altered by the presence of a strongly magnetized plasma. In particular, the GW can interact with the EMW producing excitations in the plasma, that is, Alfv\'en and magnetosonic waves. High frequency GWs produced by the NSs merger propagate through the magnetosphere of the system, interacting with the electromagnetic fields coupled with matter. This interaction (or coupling) can excite MHD waves, generating higher harmonics, such as electromagnetic radiation in the radio frequency band \cite{75,76}.

On the other hand, \cite{77} showed that a GW generated by a magnetar might well be the source of the energy released in a giant flare and may be even in a short GRB, through the absorption by the plasma of the GW energy. The authors assumed a NS magnetic field $B\sim 10^{16}\,{\rm G}$, background density $\sim 10^{-14}\; \rm{g\, cm}^{-3}$, an adiabatic index $\Gamma=4/3$, and GW frequency reaching $\sim 5\,{\rm kHz}$.

The GW-EMW coupling is more efficient if the waves are in coherence, i.e., if the frequencies satisfy some matching conditions and the relative wave phase remains unaltered for a long time \cite{78}. In fact, this is a resonant condition \cite{65,74,79}. Coherent interaction only requires that the frequencies coincide and that they have identical phase velocity. According to \cite{78}, considering GW-EMW interaction in a ``medium'', the occurrence of such resonances is more rare because the GW velocity is equal to the light speed but, the velocity of a MHD wave depends on the Alfv\'en velocity. Thus, it decreases with the mechanical pressure and increases with the magnetic pressure.

The condition for coherence to occur can be established in strongly magnetized plasmas with pressure $p$ and magnetic field $B$ if $2\mu_0p/B^2\ll 1$. In this paper, we study the interaction between GWs and EMWs in a strongly magnetized plasma, modeled by the general relativistic magnetohydrodynamics (GRMHD) equations before merger, and considering the magnetosphere density proportional to the Goldreich-Julian corotation density \cite{80}. When the GW and EMW frequencies are very close, the electromagnetic energy reaches high values.

In particular, our aim is to present the set of GRMHD equations describing the interaction of GWs with the magnetosphere of NSs. This interaction produces EMWs in the magnetized plasma during the binary inspiral phase.

We present a semianalytical formulation to describe the interaction of the $+,\times$-polarizations with the MHD wave modes. From this coupling, energy can be transferred from GWs to EMWs. Additionally, we have shown that the so-called magnetosonic mode, satisfying the coherence condition between GWs and EMWs, could reach energies close to those inferred for GW170817. We discuss what parameters and physical conditions could contribute to this.

The paper is organized as follows. In Sec. \ref{secEquations}, we present the linearized general relativity (GR) equations and the wave solution for the inspiral of compact binaries. The current densities that are generated by the GWs are presented with the linearized Maxwell equations, using the 1+3 orthonormal frame (ONF) formalism. Moreover, we show that the approximation of ideal MHD theory remains valid during the coalescence of the binary system. In Sec. \ref{secCoupling}, we demonstrate by the closed  GRMHD set equations how the interaction between the plasma and the GWs happens. With the dispersion relation derived by the Fourier Transform (FT), we show how to obtain the solutions for the equations that describe the coupling of the $+$ and $\times$ GW polarizations with the EMWs using the comovel system. For the sake of completeness, we also present these solutions in the rest frame. In Sec. \ref{secEnergy}, we discuss how much energy can be associated with
these processes and we apply our formalism in two examples, the simulation developed by \cite{64} and the source GW170817. In Sec. \ref{secRefIndex}, we study the refractive index of these systems. The Poynting vector of the MHD waves is discussed in Sec. \ref{secPoyntingVector}. Finally, our results are summarized and further discussed in Sec. \ref{remarks}, where we also discuss the relevance of our formalism for studies involving sGRBs.


\section{\label{secEquations}General Relativistic Magnetohydrodynamics Equations}

\subsection{Gravitational waves}

The GWs are described as ripples that propagate with speed of light on the space-time. Their solutions can be obtained from Einstein's field equations \cite{81}. The derivation of the wave equation solutions can be simplified, neglecting, in large scale, the curved structure of space-time by the distribution of matter, e.g., GWs of either astronomical objects with intense gravitational fields or catastrophic events can be calculated far from source, as weak ripples on the flat background \cite{82}.

Einstein's full equations, Eq.~(\ref{eqnFieldRG}), depend on the space-time metric tensor ($g_{ab}$) and the momentum-energy tensor ($T_{ab}$). The first one results in the Ricci tensor ($R_{ab}$) and scalar curvature ($R$), configuring in the geometric part of Einstein's field equations. Last, it is responsible for curvature of space-time defined by the metric $g_{ab}$ \cite{83}. $G$ and $c$ are the gravitational constant and speed of light, respectively.

\begin{equation}
\label{eqnFieldRG}
R_{ab}-\frac{1}{2}g_{ab}R=\frac{8\pi G}{c^4}T_{ab} \ .
\end{equation}

In the linearized theory, the metric tensor of the perturbed space-time differs from the Minkowski (flat) metric. The metric $g_{ab}$ usually splits as  $g_{ab}=\eta_{ab}+h_{ab}$, with $|h_{ab}|\ll 1$ \cite{84} and $\eta_{ab}=(-1,1,1,1)$. With negligible second order terms in $h_{ab}$ and the conservation of energy-momentum ($\partial_{a}T_{ab}=0$), the field equations reduce to the set of nonlinear equations \cite{85}

\begin{equation}
\label{eqnWave}
    \square\bar{h}_{ab}=-\frac{16\pi G}{c^4}T_{ab} \ .
\end{equation}

The general solution of Eq.~(\ref{eqnWave}) is given by Green functions, applying the transverse traceless (TT) gauge. The system stays in the comoving frame and the $h_{ab}$ tensor is projected on the perpendicular wave propagation direction \cite{86}. The perturbation $h_{ab}$ does not ``compress" nor ``stretch" the space-time elements, but induces a ``strain", keeping its volume.

Therefore, the GWs are purely transverse with two polarization modes: $h_{+}$ and $h_{\times}$. These modes differ by $\pi/4$ rotation around the propagation axis, which satisfies the quadruple nature of the gravitational field \cite{83,87}. In general, the GW is the superposition of these two polarizations.

After a multipole expansion and for a distant observer, the perturbation $h_{ab}$ can be written in terms of quadrupole momentum tensor that depends on the matter density. Thus, the gravitational emission is not isotropic and the symmetric motions of the system do not emit any gravitational radiation. The amplitude $h^{TT}_{\alpha\beta}$ in terms of the quadrupole momentum tensor is

\begin{equation}
\label{eqnOndaQuad}
    h^{TT}_{\alpha\beta}(t,\vec{r}) =\frac{2G}{r\,c^4}\,\ddot{Q}^{TT}_{\alpha\beta}(t-r/c) \ ,
\end{equation}
where 
\begin{equation}
\label{eqnQuadTensor}
    Q^{\alpha\beta}(t) \equiv \int \rho(t,\vec{r})\bigg(x^{\alpha}x^{\beta}-\frac{1}{3}r^2\delta^{\alpha\beta}\bigg)d^3x \ .
\end{equation}

For a binary system before coalescence, with masses $m_{1}$ and $m_{2}$ and orbital displacement in the plane $x-y$, the equations of motion are harmonic oscillations and they are given by

\begin{subequations}
\label{eqnSistemCoord}
    \begin{align}
    x_{0} &=R\cos(\omega_{s}t+\pi/2) \ , \\
    y_{0} &=R\,\mathrm{sen}\,(\omega_{s}t+\pi/2) \ , \\
    z_{0} &=0 \ .
    \end{align}
\end{subequations}

Initially, the loss of energy of the system due to the GWs can be neglected. Thus, in the Newtonian approach, the system becomes as the one-body with equal-mass case where the reduced mass is $\mu=m_{1}m_{2}/(m_{1}+m_{2})$. The Newtonian orbital frequency is given by $\omega^2_{s}=G(m_{1}+m_{2})/R^3$, where $R$ is the orbital radius \cite{88}. From these conditions, it is possible to calculate the quadrupole momentum tensor, Eq.~(\ref{eqnQuadTensor}), hence, the perturbation $h_{ab}$ for a binary system. The GW amplitude, in the Fourier space, released by the binary system in the inspiral phase is given by \cite{88}

\begin{subequations}
\begin{align}
h_{+}(f_{GW}) &= A e^{i\psi_{+}(f_{GW})}\frac{c}{r}\bigg(\frac{G M_{c}}{c^3}\bigg)^{5/6}\frac{1}{f_{GW}^{7/6}}\bigg(\frac{1+\cos^2\iota}{2}\bigg) \label{eqnOGBinFourier:x} \ , 
\\
h_{\times}(f_{GW}) &= A e^{i\psi_{\times}(f_{GW})}\frac{c}{r}\bigg(\frac{G M_{c}}{c^3}\bigg)^{5/6}\frac{1}{f_{GW}^{7/6}}\cos^2\iota \label{eqnOGBinFourier:+}\ ,
\end{align}
\end{subequations}
where the chirp mass is $M_{c}={(m_{1}m_{2})^{3/5}}/{(m_{1}+m_{2})^{1/5}}$, $A={\pi^{-2/3}}({5}/{24})^{1/2}$,  $\psi_{+}(f_{GW})$, and $\psi_{\times}(f_{GW})$ are the phases of $+, \times$ given by

\begin{align}
\label{eqnGWPhases}
\psi_{+}(f_{GW})&=2\pi f_{GW}\, t_{coal} - \phi_{c} - \frac{\pi}{4} + \frac{3}{4}\bigg(\frac{G M_{c}}{c^3}8\pi f_{GW}\bigg)^{-5/3} \ , \\
\psi_{\times}(f_{GW}) &= \psi_{+}(f_{GW}) + \pi/2 \ .
\end{align}

Equations~(\ref{eqnOGBinFourier:x}) and (\ref{eqnOGBinFourier:+}) are the spectral amplitude densities. They have units of amplitude intensity over frequency.

The GW frequency increases with time ($t$) until the system reaches the merger at $t_{coal}$. In terms of the parameter $\tau=t_{coal}-t$, we have

\begin{equation}
\label{eqnFreqOG_1}
    f_{GW}(\tau) = \frac{1}{\pi}\bigg(\frac{5}{256}\frac{1}{\tau}\bigg)^{3/8}\bigg(\frac{G M_{c}}{c^3}\bigg)^{-5/8} \ ,
\end{equation}

The GW frequency depends on the chirp mass ($M_c$) of the binary system, the distance from the observer ($r$) and the $\iota$ angle with the normal of the orbital system. While the system coalesces, the orbital radius decreases with the time $\tau$ as

\begin{equation}
    \label{eqnROG_1}
    R(\tau)=R_{0}\Big(\frac{\tau}{\tau_{0}}\Big)^{1/4} \ ,
\end{equation}
where $R_{0}$ is the initial orbital radius at $t_{0}$, thereby $\tau_{0}=t_{coal}-t_{0}$. 

With the equations for $f_{GW}$ and $R$, Eqs.~(\ref{eqnFreqOG_1}) and~(\ref{eqnROG_1}), respectively, the orbital velocity of the stars in terms of the GW frequency is

\begin{equation}
    \label{eqnVelOrb}
    v = [\pi G (m_1+m_2)]^{1/3} f_{GW}^{1/3} \ .
\end{equation}

When the stars are very close, the gravitational field is strong and so there are important consequences on the dynamics of the binary system. In fact, there is a minimum distance between the stars able of keeping the orbit circularly stable \cite{89}. It is called inner-most stable circular orbit (ISCO). On Schwarzschild coordinates, we have

\begin{equation}
    \label{eqnR_ISCO}
    R_{(ISCO)}=\frac{6Gm}{c^2} \,
\end{equation}
where $m=m_{1}+m_{2}$.

Therefore, the adiabatic inspiral phase occurs at distances $R \gtrsim R_{(ISCO)}$. Close to $R_{(ISCO)}$, the dynamics are dominated by strong field effects and the two stars plunge toward each other. The inspiral phase ends when the merger begins. The GW frequency at the ISCO is

\begin{equation}
    \label{eqnF_isco}
    f_{GW_{(ISCO)}} =\frac{1}{3\sqrt{6}}\frac{c^3}{2\pi G m} \ .
\end{equation}

The emission of GWs removes energy from the system. In this way, $R$ must decrease with time while the $f_{GW}$ increases. The power radiation is then

\begin{equation}
    \label{eqnP}
    P=\frac{32}{5}\frac{c^5}{G}\bigg(\frac{GM_{c}\pi f_{GW}}{c^3}\bigg)^{10/3}\ .
\end{equation}

In Newtonian formalism, the total energy released is given by

\begin{equation}
    \label{eqnGW_Energy}
    \Delta E_{rad} \sim \frac{\pi}{2G}(GM_{c})^{5/3}\Big( f_{GW_{(ISCO)}}\Big)^{2/3} \ . 
\end{equation}

Considering the system GW170817 with total mass $m=2.73 M_{\odot}$, chirp mass $M_c=1.188 M_{\odot}$, the frequency is $f_{GW_{ISCO}} \approx 1.61$ kHz, and $R_{(ISCO)}\approx2.42\times 10^4$ m. The power radiation evolves like $P\simeq 7.49 \times 10^{36} f_{GW}^{10/3}$ and the energy released is about $\Delta E_{rad}\sim 1.01\times 10^{46}$ J. In these conditions, Eq.~(\ref{eqnVelOrb}) produces $\beta=v/c\approx0.406$ in the inner-most stable circular orbit. 

\subsection{Electromagnetic fields}

We want to see how the GWs induce small perturbations in the plasma around the stars that form the binary system. The plasma is in the space-time described by the metric $g_{ab}=\eta_{ab}+h_{ab}$. Moreover, the plasma proprieties are averages over the fluid elements \cite{90,91}. In the MHD formalism, the equations describing the magnetized plasma come from the fluid and electromagnetic equations. We work with the Gaussian units with $c=1$ (explicitly shown otherwise, for convenience).

The Maxwell's equations in terms of the electromagnetic field tensors $F^{ab}$ (Maxwell tensor), its \textit{dual} $\mathfrak{F}^{ab}$, and the $4$-current density, $j^{b} = (\rho, \vec{j})$, are given by \cite{92}

\begin{subequations}
\begin{align}
\label{eqnMaxwellCovariant}
    \nabla_{b}F^{ab} &= 4\pi j^{a}\ ,\\
    \nabla_{b}\mathfrak{F}^{ab} &=0 \ .
\end{align}
\end{subequations}

In the 3+1 formalism \cite{93}, the Maxwell equations can be described in the comovel frame, whereas the vector basis of orthonormal tetrad is written in terms of the GW amplitude \cite{75}

\begin{align}
\label{eqnTetrada}
\hat{e}_{0} &=\bigg(\frac{\partial}{\partial t},0,0,0\bigg)  \ , \nonumber \\
\hat{e}_{1} &=\Big(0,\Big[1-\frac{h_{+}}{2}\Big]\frac{\partial}{\partial x}, \frac{-h_{\times}}{2}\frac{\partial}{\partial y},0\Big) \ , \nonumber \\
\hat{e}_{2} &=\Big(0,\frac{-h_{\times}}{2}\frac{\partial}{\partial x},\Big[1+\frac{h_{+}}{2}\Big]\frac{\partial}{\partial y},0\Big) \ , \nonumber \\
\hat{e}_{3} &=\Big(0,0,0,\frac{\partial}{\partial z}\Big) \ ,
\end{align}
and the metric tensor $g^{ab}$ is described by

\begin{equation}
\label{eqnMetrica}
g^{ab}(t,z) =
\begin{pmatrix}
-1  &0          &0      &0 \\ 
0   &1+h_{+}    &h_{\times}  &0 \\ 
0   &h_{\times}      &1-h_{+} &0 \\ 
0   &0          &0      &1 
\end{pmatrix} \ .
\end{equation}

The linearized Maxwell equations coupled to the gravitational perturbations, in the specified tetrad, are \cite{76}

\begin{align}
\label{eqnMaxwell}
\nabla \times \vec{E}^{(1)} + \frac{\partial \vec{B}^{(1)}}{\partial t} &= - \vec{j}_{B}^{(1)} \ ,\nonumber\\
\nabla \times \vec{B}^{(1)}-\frac{\partial \vec{E}^{(1)}}{\partial t} &= 4\pi \vec{j}^{(1)}+\vec{j}_{E}^{(1)} \ ,\nonumber\\
\nabla \cdot \vec{E}^{(1)} &= 4 \pi \rho^{(1)} \ ,\nonumber \\
\nabla \cdot \vec{B}^{(1)} &= 0 \ .
\end{align}

In the plasma rest frame and considering a collisionless plasma with no dissipative effects and conductivity $\sigma \rightarrow \infty$, then the electric field ($\vec{E}^{(0)}$), the plasma velocity ($\vec{v}^{(0)}$), the current density ($\vec{j}^{(0)}$), and the charged matter density ($\rho^{(0)}$) all disappear.

We take the covariant derivative $\nabla_{c}F^{ab} = \partial_{c}F^{ab}+\Gamma^{a}_{\phantom{a}dc}F^{db} + \Gamma^{b}_{\phantom{b}cd}F^{ad}$ \cite{87} and we disregard terms of order two or higher $(\mathcal{O}(h_{+,\times}^2) \approx 0)$. Here, the superscript $(0)$ corresponds to the background variables and the notation $(1)$ corresponds to the perturbed variables.

The current densities induced ``gravitationally'', $\vec{j}_{B}$ and $\vec{j}_{E}$ in Eq.~(\ref{eqnMaxwell}), are calculated by 

\begin{subequations}
\label{eqnCorrente}
\begin{align}
\vec{j}_{B}^{(1)} &= -\frac{B^{(0)}_{x}}{2}\frac{\partial}{\partial t} \begin{pmatrix}
                                                                h_{+}       \\ 
                                                                h_{\times}  \\ 
                                                                0   
                                                                \end{pmatrix} \ ,\\
\vec{j}_{E}^{(1)} &= \frac{B^{(0)}_{x}}{2}\frac{\partial}{\partial z} \begin{pmatrix}
                                                                h_{\times} \\ 
                                                                -h_{+}       \\ 
                                                                0   
                                                                \end{pmatrix} \ .
\end{align}
\end{subequations}

We note that in \cite{76}, the expression $\vec{j}_{E}^{\,(1)}$ has a different signal from that obtained here. However, the physical interpretation of this parameter obtained by \cite{76} is the same as that obtained here. The effect of the GWs is to induce small perturbations in all quantities of the plasma. Therefore, all equations are linearized around the unperturbed state. Important to note that the functions $h_{+,\times}(z,t)$ are general, their waveforms depend on the source of gravitational radiation.

\subsection{Plasma equations}
\label{subsec_plasma}

In the ideal MHD formalism, the plasma is described as a unique fluid, collisionless, without viscosity (and so the pressure tensor $\mathbb{P} = p\mathbb{I}$, where $\mathbb{I}$ is the identity matrix), and no heat dissipation (that is, the heat flux $\vec{Q}=0$) \cite{90}.

The gravitational terms are conservative, besides they are neglected in comparison with the electromagnetic terms. Without dissipative effects, the resistivity of the plasma is insignificant (i.e., the conductivity $\sigma \rightarrow \infty$). For a plasma with infinite conductivity, the particles quickly restore the neutrality condition causing the charge $\rho(\vec{r},t)$ and the current ($\vec{j}$) densities to disappear \cite{91}. Therefore, the system is adiabatic and the energy conservation condition produces   

\begin{equation}
\label{eqnAdiabatica}
p^{(0)} = k(\rho_{m}^{(0)})^{\Gamma} \ ,
\end{equation}
where $\Gamma$ is the polytropic index ($4/5 \leqslant \Gamma \leqslant 5/3$).

By the first law of thermodynamics $dU =dQ - pdV$ \cite{94}, where $U$ is the internal energy per unit mass, $p$ is the plasma pressure, and $V$ is the specific volume per unit mass ($V=1/\rho_{m}$). With Eq.~(\ref{eqnAdiabatica}), the internal energy is

\begin{equation}
\label{eqn1termo}
    U^{(0)} = \frac{p^{(0)}}{\rho_{m}^{(0)}(\Gamma-1)} \ .
\end{equation}
 
Hence, the total relativistic energy-matter of the plasma with respect to the $4$-velocity is given by (the speed of light $c$ is shown for convenience)

\begin{equation}
\label{eqnDensidade}
\mu^{(0)} = \rho_{m}(c^2+U^{(0)}) = \rho_{m}^{(0)} c^2 + \frac{p^{(0)}}{\Gamma-1} \ .
\end{equation}

The relativistic enthalpy (with unit $\rm{N m^{-2}}$) is given by \cite{54,94}

\begin{equation}
\label{eqnEntalpia}
w^{(0)} = \mu^{(0)} + p^{(0)} \ ,
\end{equation}
such that $\mu^{(0)}$ and $p^{(0)}$ are considered mechanical pressures.

Combining Eqs.~(\ref{eqnAdiabatica}),~(\ref{eqnDensidade}), and~(\ref{eqnEntalpia}), the proper relativistic sound velocity can be calculated by

\begin{equation}
\label{eqnSom}
c_{s}^2 = \frac{\partial p}{\partial \mu}\bigg|_{ad} = \frac{\Gamma p^{(0)}}{w^{(0)}} \ .
\end{equation}

The pressure gradient can be written as $\nabla p^{(1)} = c^{2}_{s}\nabla \mu^{(1)}$. The enthalpy, in Eq.~(\ref{eqnEntalpia}), results in the matter density $\mu^{(0)}$, when the plasma pressure is neglected or when the plasma is in the cold plasma approximation. 

The conservation of energy-momentum is determined by

\begin{equation}
\nabla_{b}T^{ab} = \nabla_{b}\Big[(\mu + p)u^au^b+pg^{ab} + \frac{1}{4\pi}\big(F^{a}_{\phantom{a}c}F^{bc} - \frac{1}{4\pi}F^{cd}F_{cd}\big)\Big] \ .
\end{equation}

The conservation of matter density in the rest frame is

\begin{equation}
\label{eqnMatterDensity}
\frac{\partial \rho_{m}^{(1)}}{\partial t} + \rho_{m}^{(0)}\nabla \cdot \vec{v}^{(1)} = 0  \ .
\end{equation}

In the comoving frame, the electric field ($E^{(0)}$), the velocity of the plasma ($\vec{v}^{\,(0)}$), and the current density ($\vec{j}^{\,(0)}$) disappear. Therefore, the conservation equation for the energy-matter density, when the terms of higher order are negligible, is given by

\begin{equation}
\label{eqnDensidadeComovel}
\frac{\partial p^{(1)}}{\partial t} + \Gamma p^{(0)} \nabla \cdot \vec{v}^{(1)} = 0 \ .
\end{equation}

In the nonrelativistic limit when the internal energy is negligible with respect to the rest-mass energy ($p\ll\mu$), Eq.~(\ref{eqnDensidadeComovel}) reduces to Eq.~(\ref{eqnMatterDensity}).

The momentum conservation equation in the comovel frame can be determined by

\begin{equation}
\label{eqnMomento}
\big(\mu^{(0)} + p^{(0)}\big)\frac{\partial \vec{v}^{(1)}}{\partial t} + \nabla p^{(1)} = \vec{j}^{(1)}\times\vec{B}^{(0)} \ .
\end{equation}

To complete the set of equations describing the electromagnetic field in the ideal MHD approximation, it is necessary to calculate the Ohm law. That is,

\begin{align}
\frac{m_{e}}{n_{n}e^2} \frac{\partial \vec{j}}{\partial t} - \frac{1}{n_{n}e}\nabla \cdot \mathbb{P}_{e}= \vec{E} + \vec{v}\times\vec{B}-\frac{1}{n_{n}e}[\vec{j}\times\vec{B}]-\frac{1}{\sigma} \vec{j} \ .
\end{align}

Considering the very conductive plasma ($\sigma\rightarrow\infty$), the quasineutrality condition ($\vec{E}^{(0)}=\vec{j}^{(0)}=\rho^{(0)}=0$), collisionless plasma, and without Joule effect ($\vec{j}\cdot\vec{E} =0$), then the Hall effect $(\vec{j}\times\vec{B})$ and the $\partial \vec{j}/\partial t$ terms disappear in the generalized Ohm's Law. Thus, as in the comoving frame the plasma velocity is $\vec{v}^{(0)}=0$, we can write the generalized Ohm's law as

\begin{equation}
    \label{eqnOhm}
    \vec{E}^{(1)} = -\vec{v}^{(1)}\times\vec{B}^{(0)} \ .
\end{equation}

Thereby, we have the closed set of the variables describing the strongly magnetized plasma, in the ideal MHD theory, and that is interacting with the GWs emitted by BNSs.

\section{Coupling GWs to EMWs}
\label{secCoupling}

\subsection{GRMHD equations}
\label{subsec_GRMHD}

The set of partial differential equations for the 16 variables $\vec{B}, \vec{E}, \vec{j}, \rho_{m}, \vec{v}, \mu, \rho, p$ --- Eqs.~(\ref{eqnMaxwell}, \ref{eqnAdiabatica}, \ref{eqnMatterDensity}, \ref{eqnDensidadeComovel}, \ref{eqnMomento}, \ref{eqnOhm}) --- describing the relativistic and strongly magnetized MHD plasma that is coupled to the GWs generated by the evolving binary system, in a Gaussian system with $c=1$, is \cite{68,76}

\begin{align}
\nabla \times \vec{E}^{(1)} + \frac{\partial \vec{B}^{(1)}}{\partial t} &= - \vec{j}_{B}^{(1)} \label{eqnGRM:1}\ ,\\
\nabla \times \vec{B}^{(1)}-\frac{\partial \vec{E}^{(1)}}{\partial t} &= 4\pi \vec{j}^{(1)}+\vec{j}_{E}^{(1)} \label{eqnGRM:2}\ ,\\
\nabla \cdot \vec{E}^{(1)} &= 4 \pi \rho^{(1)} \label{eqnGRM:3}\ ,\\
\nabla \cdot \vec{B}^{(1)} &= 0  \label{eqnGRM:4}\ ,
\end{align}

\begin{align}
\frac{\partial p^{(1)}}{\partial t} + \Gamma p^{(0)} \nabla \cdot \vec{v}^{(1)} &= 0 \label{eqnGRM:5}\ ,\\
\nabla p^{(1)} &= c^{2}_{s}\nabla \mu^{(1)} \label{eqnGRM:6}\ , \\
\frac{\partial \rho_{m}^{(1)}}{\partial t} + \rho_{m}^{(0)}\nabla \cdot \vec{v}^{(1)} &= 0 \label{eqnGRM:7}\ , \\
\big(\mu^{(0)} + p^{(0)}\big)\frac{\partial \vec{v}^{(1)}}{\partial t} + \nabla p^{(1)} &= \vec{j}^{(1)}\times\vec{B}^{(0)} \label{eqnGRM:8}\ , \\
\vec{E}^{(1)} &= -\vec{v}^{(1)}\times\vec{B}^{(0)} \label{eqnGRM:9}\ ,
\end{align}
where the gravitationally induced current densities are

\begin{align}
\vec{j}_{B}^{(1)} &= -\frac{B^{(0)}_{x}}{2}\frac{\partial}{\partial t}\begin{pmatrix}
                                                                h_{+}       \\ 
                                                                h_{\times}  \\ 
                                                                0   
                                                               \end{pmatrix} \ ,
\end{align}

\begin{align}                                                           
\vec{j}_{E}^{(1)} &= \frac{B^{(0)}_{x}}{2}\frac{\partial}{\partial z} \begin{pmatrix}
                                                                h_{\times} \\ 
                                                                -h_{+}       \\ 
                                                                0   
                                                                \end{pmatrix} \ . \\
                                                                \nonumber
\end{align}

The matter density and relativistic enthalpy are, respectively,
                                                             
\begin{align}
\mu^{(0)} &= \rho_{m}^{(0)} + \frac{p^{(0)}}{\Gamma-1} \ ,\\
w^{(0)} &= \mu^{(0)} + p^{(0)} \ .
\end{align}

\subsection{Dispersion relation}

\begin{figure}[!h]
\includegraphics[scale=0.65]{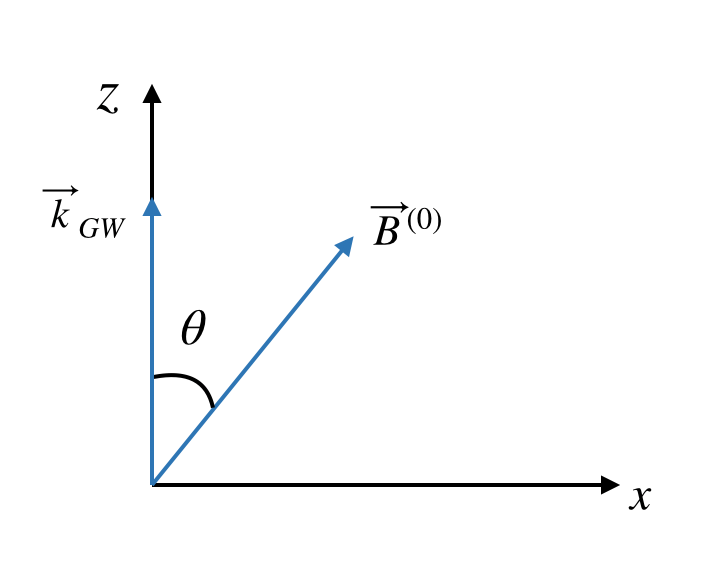}
\caption{\label{fig:xz_plane} The background magnetic field forms the  $\theta$ angle with the GW vector that is in the $z$-direction. There is no loss of generality with this choice.}
\end{figure}

The differential equations are calculated for the strongly magnetized plasma with background magnetic field $B^{(0)}(x,z)$, oriented in the $x-z$ plane and forming a $\theta$ angle with the $z$-axis. The GW amplitudes $h_{\times,+}(z,t)$ propagate along the $z$-axis (see Fig. \ref{fig:xz_plane}).

We are interested in the scenario where the GWs and the MHD waves are in coherence, i.e., when they have the same propagation direction, frequency, and the phase difference remains constant. This means that they must have very close phase velocities. In this way, the waves may interact through constructive or destructive interference and so they can exchange energy.

Taking the second time derivative of Eq.~(\ref{eqnGRM:8}), eliminating $p^{(1)}$ by the use of Eq.~(\ref{eqnGRM:5}) and $\vec{B}^{\,(1)}$ with the aid of Eqs.~(\ref{eqnGRM:1}-\ref{eqnGRM:2}), then using Eq.~(\ref{eqnGRM:3}) to eliminate $\vec{E}^{\,(1)}$, after that using the expressions for $\vec{j}^{\,(1)}_E$ and $\vec{j}^{\,(1)}_ B$, we can obtain the plasma perturbation wave equation in terms of the velocity $\vec{v}^{\,(1)}(z,t)$ of the perturbations. That is,

\begin{widetext}
\begin{align}
\label{eqnAcoplamento}
\bigg(\frac{\partial^2}{\partial t^2} - u_{m}^2 \nabla \nabla \cdot \bigg) \vec{v}^{(1)}&-\bigg[\vec{u}_{A}\frac{\partial^2}{\partial t^2}-(\vec{u}_{A}\cdot\nabla)\nabla\bigg]\vec{v}^{(1)}\cdot\vec{u}_{A}\nonumber \\ 
&=(\vec{u}_{A}\cdot\nabla)^2\vec{v}^{(1)}-\vec{u}_{A}(\vec{u}_{A}\cdot\nabla)(\nabla\cdot\vec{v}^{(1)}) +\frac{1}{\sqrt{4\pi\omega_{tot}}}\bigg[\nabla(\vec{u}_{A}\cdot\vec{j}_{B}^{(1)})-\frac{\partial}{\partial t}(\vec{j}_{E}^{(1)}\times\vec{u}_{A})-(\vec{u}_{A}\cdot\nabla)\vec{j}_{B}^{(1)}\bigg] \ ,
\end{align}
\end{widetext}
where the relativistic Alfv\'en velocities for a noncompressional (shear Alfv\'en) wave, $u_{A}$, and for a magnetoacoustic wave, $u_{m}$ (with magnetic enthalpy, that corresponds to mechanical and magnetic pressures, $\omega_{tot} = w^{(0)} + |\vec{B}^{(0)}|^2/(4\pi$)), are, respectively, defined as

\begin{align}
u^2_{A} = \frac{|\vec{B}^{(0)}|^2}{4\pi\omega_{tot}} \label{eqnVelAlfven} \ ,\\
u^2_{m} = \frac{\Gamma p^{(0)}}{\omega_{tot}} + \frac{|\vec{B}^{(0)}|^2}{4\pi\omega_{tot}} \ . \label{eqnVelMagneto}
\end{align}

Equation~(\ref{eqnAcoplamento}) is calculated algebraically in the Fourier space. By applying the Fourier transform with respect to time and spatial parts, where $\omega$ and $k$ are the frequency and the wave vector of the MHD modes, respectively, and using the following definitions:

\begin{align}
\vec{B}^{(0)}&=B_{x}^{(0)}\hat{x}+B_{z}^{(0)}\hat{z} \ , \\
\vec{u}_{A} &= u_{A_{\perp}}\hat{x} + u_{A_{\parallel}}\hat{z} \ , \\
\vec{v}^{(1)}(z,t) &= v_{x}^{(1)}(z,t)\hat{x} + v_{y}^{(1)}(z,t)\hat{y}+v_{z}^{(1)}(z,t)\hat{z} \ ,
\end{align}
we can obtain the dispersion relation

\begin{widetext}
\begin{align}
\begin{pmatrix}
\omega^2(1-u^2_{A\perp})-k^2u^2_{A\parallel} &0                             &-(\omega^2-k^2)u_{A\parallel}u_{A\perp} \\ 
0                                            &\omega^2-k^2u_{A\parallel}^2    &0 \\ 
-(\omega^2-k^2)u_{A\parallel}u_{A\perp}      &0                     &\omega^2(1-u^2_{A\parallel})-k^2(u_{m}^2-u^2_{A\parallel}) 
\end{pmatrix} \vec{v}^{(1)} = -u_{A\perp}\omega k\begin{pmatrix}
                                                            u_{A\parallel}h_{+}       \\
                                                            u_{A\parallel}h_{\times}  \\
                                                            -u_{A\perp}h_{+}   
                                                            \end{pmatrix} \ .
\label{eqnDispersaoTermos}
\end{align}
\end{widetext}

Equation~(\ref{eqnDispersaoTermos}) can be represented as $D\,\vec{v}^{\,(1)}=\textbf{J}^{\,(1)}_{GW}$. The left side is purely magnetohydrodynamics. It describes the behavior of the plasma through the Alfv\'en and magnetosonic modes. The right side is purely gravitational and this term excites the plasma parameters and its respective modes.

The amplitudes $h_{+,\times}$ on the Fourier space are the source terms and they depend on the evolutionary stage of the binary system. In \cite{76}, it was used an impulse function as $\delta(\omega-\omega_{GW})\propto\mathcal{F}\{h_{+,\times}e^{i\omega_{GW}(z-t)}\}(\omega)$ to consider only the instant immediately before the merger. In \cite{68}, it was used a waveform that reproduces the inspiral phase until instants before the merger as represented by Eqs.~(\ref{eqnOGBinFourier:x})-(\ref{eqnOGBinFourier:+}).

The homogeneous solution of Eq.~(\ref{eqnDispersaoTermos}) is obtained when the gravitational source is turned off ($h_{+, \times}=0$). Thus, it is necessary  to calculate the determinant of the $D^{-1}$-matrix, where

\begin{widetext}

\begin{equation}
\label{eqnDInversa}
D^{-1} =\begin{pmatrix}
\frac{\omega^2(u^2_{A\parallel}-1)-k^2(u_{m}^2-u^2_{A\parallel})}{\Lambda}  &0 &\frac{(\omega^2-k^2)u_{A\parallel}u_{A\perp}}{\Lambda} \\ 
0                                                                            &\frac{1}{\omega^2-k^2u^2_{A\parallel}}    &0 \\ 
\frac{(\omega^2-k^2)u_{A\parallel}u_{A\perp}}{\Lambda}       &0     &\frac{\omega^2(1-u^2_{A\perp})-k^2u^2_{A\parallel}}{\Lambda}
\end{pmatrix} \ .
\end{equation}

\end{widetext}

Denominating the determinant of $D^{-1}$ as $\Lambda(\omega,k)$, we have

\begin{align}
\Lambda(\omega,k) &= u^2_{A_{\parallel}}c^2_{s}k^4 \\ \nonumber
&+ \omega^2\big[(u^2_{A_{\perp}}-1)u^2_{m} + u^2_{A_{\parallel}}u^2_{A}\big]k^2 - \omega^4(u^2_{A}-1) \ ,
\end{align}

or in terms of the eigenvalues

\begin{equation}
\Lambda(k,\omega)=(1-u_{A}^2)(\omega^2-k^2u_{s}^2)(\omega^2-k^2u_{f}^2) \ .
\end{equation}

The homogeneous solutions are

\begin{align}
\omega &= \pm k_{A}u_{A\parallel}\ , \label{eqnModo:1}
\\
\omega &= \pm \frac{k_{s,f}}{\sqrt{2}}\sqrt{\Big(u_{m}^2+c^2_{s}\frac{u_{A\parallel}^2}{1-u_{A}^2}\Big)\sqrt{1\pm\sqrt{(1-\sigma)}}} \label{eqnModo:2} \ .
\end{align}

The auxiliary parameter $\sigma(\theta)$ in Eq.~(\ref{eqnModo:2}) is given by

\begin{equation}
\label{eqnSigma}
\sigma(\theta)\equiv {4c_{s}^2\frac{u_{A\parallel}^2}{1-u_{A}^2}} {\left(u_{m}^2+c_{s}^2\frac{u_{A\parallel}^2}{1-u^2_{A}}\right)^{-2}},
\end{equation}
where our expression differs by the factor $1/(1-u_A^2)$ of that obtained by \cite{76}.

As expected, the perturbations produce shear Alfv\'en waves (AW) and compressional magnetosonic waves (MSW). The solutions given in Eqs.~(\ref{eqnModo:1}) and (\ref{eqnModo:2}) represent 6 equations describing the excitation modes of the plasma. The negative sign inside the square root of Eq.~(\ref{eqnModo:2}) refers to the slow MSW with phase velocity $u_{s}=\omega/k_{s}$. The positive sign refers to the fast MSW with phase velocity $u_{f}=\omega/k_{f}$. The equations for $u_{f}$ and $u_{s}$ are represented by

\begin{equation}
\label{eqnVelocidadeFase}
u^2_{f,s} = c_{s}\frac{u_{A_{\parallel}}}{\sqrt{1-u_{A}^2}}\bigg(\frac{1 \pm \sqrt{1-\sigma}}{\sqrt{\sigma}}\bigg) \ .
\end{equation}

Seeing that the $D^{-1}$-matrix is inverse, the solutions for the nonhomogeneous linear system are given by $v_{\alpha}^{\,(1)} = \big(D^{-1}\vec{J}^{\,(1)}_{GW}\big)_{\alpha}$.

We can see that the AW mode is excited by GW polarization $h_{\times}$, while the MSW mode couples to the GW polarization $h_{+}$.  

\subsection{Comoving frame}
\label{subsec_CoFrame}

The equations presented above are written in the proper frame. Hence, the results obtained are related to the frames moving with the plasma velocity around of the compact binary. Considering the plasma frozen in the magnetic field lines, the plasma velocity is the same of the stars, as represented in Eq.~(\ref{eqnVelOrb}).

\subsubsection{Alfv\'en waves}
\label{subsubsec_AW}

The $D^{-1}_{yy}$ component couples to the $\times$-polarization and excites Alfv\'en waves (AWs). Calculating $v_{y} = D^{-1}_{yy}J_{GW_{yy}}^{(1)}$ with $D^{-1}$ from Eq.~(\ref{eqnDInversa}), we have

\begin{equation}
\label{eqnVelLaplaceAlfvenCoal}
v_{y}^{(1)}(k,\omega) = -h_{\times}(\omega_{GW})\frac{u_{A_{\parallel}} u_{A_{\perp}}}{\omega^2-k^2u_{A_{\parallel}}^2} \omega k \ .
\end{equation}

The perturbed velocity is along the $y$-axis, perpendicular to the background magnetic field of the binary system ($v^{\,(1)}_{y}\perp\vec{B}^{\,(0)}$). This defines the Alfv\'en velocity and its direction as $v^{\,(1)}_{y}\perp\vec{k}_{A}$. The velocity propagation of the Alfv\'en wave is high when the background magnetic field is parallel to the $z$-axis and, considering the phase velocity, we have $\omega/k_{A}=u_{A_{\parallel}}=u_{A}\cos{\theta}$. 

The velocity $v_{y}(k,\omega)$ depends explicitly on the polarization state $h_{\times}$($\omega_{GW}$) and on the MHD wave frequency. The perturbed magnetic field ($B_{y}^{\,(1)}$), perturbed electric fields ($E_{z}^{\,(1)}$ and $E_{x}^{\,(1)}$) and other plasma quantities ($j_{x}^{(1)}$, $j_{z}^{(1)}$ and $\rho^{(1)}$) are calculated by

\begin{subequations}
\label{eqnGrandezasLaplaceAlfvenCoal}
\begin{align}
-\frac{E_{z}^{(1)} (k, \omega)}{B_{x}^{(0)}} &= \frac{E_{x}^{(1)} (k, \omega)}{B_{z}^{(0)}} = - v_{y}^{(1)}(k, \omega) \ , \\
B_{y}^{(1)} (k, \omega) &=-v_{y}^{(1)}(k,\omega)\frac{B_{x}^{(0)}}{u_{A_{\parallel}}u_{A_{\perp}}}\frac{\omega^2+k^2u^2_{A_{\parallel}}}{2\omega k} \label{eqnGrandezasLaplaceAlfvenCoal:B}\ ,\\
j_{x}^{(1)} (k, \omega) &=-\frac{i\omega}{4\pi}\frac{1-u^2_{A_{\parallel}}}{u^2_{A_{\parallel}}}E_{x}^{(1)}(k,\omega) \ ,\\
j_{z}^{(1)} (k, \omega) &= \frac{i\omega}{4\pi}E_{z}^{(1)}(k, \omega) \ ,\\
\rho^{(1)} (k,\omega) &= \frac{ik}{4\pi}E_{z}^{(1)}(k, \omega) \ .
\end{align}
\end{subequations}

\begin{figure}[ht]
\includegraphics[scale=0.85]{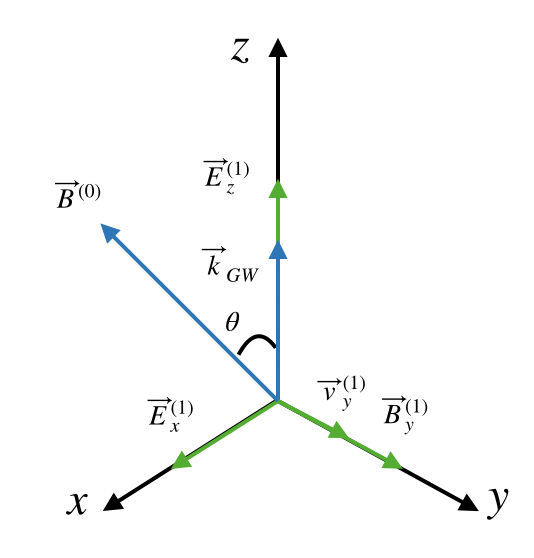}
\caption{\label{fig:xz_plane_AW} The background magnetic field is in the $x-z$ plane and does the $\theta$ angle with the GW direction. The polarization $h_{\times}$ excites the plasma, producing oscillations in the electromagnetic fields ($E^{(1)}_z, E^{(1)}_x$ and $B^{(1)}_y$). The oscillations produce shear in the magnetic field lines; this is the Alfv\'en mode \cite{91}.}
\end{figure}

The directions of the perturbed electromagnetic fields and other physical parameters are shown in Fig. \ref{fig:xz_plane_AW}. The interaction is more efficient when the wave number ($k_{A}$) of the AW is parallel to the wave number ($k_{GW}$) of the GWs, or when the background magnetic field is parallel to the GW direction, that corresponds to $\theta\rightarrow0$. In this case, the Alfv\'en phase velocity is maximum while the $z$-component of the electric field ($E^{(1)}_z$) and the magnetic field $B^{(1)}_{y}$ become both null. This result is expected and it was also obtained in studies developed by \cite{73,74}. 

Because the AW is not compressional, it does not produce perturbations in the pressure $p^{(0)}$ and matter density $\rho^{(0)}_{m}$. Therefore, it makes no sense to calculate $p^{(1)}$ and $\rho^{(1)}_{m}$. On the other hand, the Alfv\'en waves change the neutrality state of the plasma and so producing current densities, $j_{x}^{(1)}$ and $j_{z}^{(1)}$, which are modified as the coalescence evolves. Moreover, the AW produces shear in the background magnetic field lines. See also, using Eq.~(\ref{eqnGrandezasLaplaceAlfvenCoal:B}) that the perturbed field lines are perpendicular to $B^{(0)}_x$ during the oscillation, \cite{91}.   

\subsubsection{Magnetoacoustic waves}
\label{subsubsec_MSW}

It is expected that the $h_{+}$ and $h_{\times}$ excite slow and fast MSW in the plasma. Using $v_{x} = D^{-1}_{xx}J_{GW_{xx}}^{(1)} + D^{-1}_{xz}J_{GW_{zz}}^{(1)}$ and $v_{z} = D^{-1}_{zx}J_{GW_{xx}}^{(1)} + D^{-1}_{zz}J_{GW_{zz}}^{(1)}$ to calculate the other velocity components, we have

\begin{align}
\label{eqnVelLaplaceMagnetoCoal}
v^{(1)}_{z} (k,\omega) &= \frac{h_{+}(\omega_{GW})\omega^3ku^2_{A_{\perp}}}{(\omega^2-k^2u^2_{f})(\omega^2-k^2u^2_{s})}\ ,\\
v^{(1)}_{x} (k,\omega) &= - \frac{v_{z}(k,\omega)}{\tan\theta}\bigg[1-\frac{c^2_{s}k^2}{\omega^2(1-u_{A}^2)}\bigg] \label{eqnVelLaplaceMagnetoCoal:x}\ ,
\end{align}
where $\tan\theta=u_{A_{\perp}}/u_{A_{\parallel}}$ by definition of the Alfv\'en velocity. We observe that Eq.~(\ref{eqnVelLaplaceMagnetoCoal:x}) differs by a factor $1/(1-u_A^2)$ from that obtained by \cite{76}.

The perturbed plasma velocities for the MSW, Eqs.~(\ref{eqnVelLaplaceMagnetoCoal}) and (\ref{eqnVelLaplaceMagnetoCoal:x}), are parallel to the background magnetic field ($v^{(1)}_{x,z}\parallel \vec{B}^{(0)}$). The GW polarization $h_{+}$ is modified as the coalescence evolves, and, as a consequence, it perturbs the plasma velocity and induces oscillations in the $x-z$ plane. By the condition expressed through Eq.~(\ref{eqnGRM:4}), the perturbation of the magnetic field is orthogonal to the direction of the plasma oscillations. The perturbed magnetic field $B_x^{(1)}$ is

\begin{align}
\label{eqnBLaplaceMagnetoCoal}
\frac{B_x^{(1)}(k, \omega)}{B^{(0)}} = v_{z}^{(1)}(\omega,k)&\sin \theta - v_{x}^{(1)}(\omega, k)\cos\theta \nonumber \\ - &\frac{\omega}{\omega+k}\frac{1-u^2_{A}}{u^2_{A}}\frac{v^{(1)}_{x}(\omega, k)}{\cos\theta} \ .
\end{align}

The MSW is an electromagnetic and compressional wave because $B^{(1)}_x$ is in the plane of the background magnetic field. The density of the magnetic field lines increases with the wave propagation, and, considering that the charged matter is frozen in these lines, we see as a result the increase in both pressure and matter density. The perturbed electric field ($E_{y}^{(1)}$), the perturbed mechanical pressure ($p^{(1)}$ and $\mu^{(1)}$), and the current density ($j_{y}^{(1)}$) are calculated as

\begin{subequations}
\label{eqnGrandezasLaplaceMagnetoCoal}
\begin{align}
p^{(1)}(k, \omega) &= \frac{k}{\omega}\Gamma p^{(0)}v_{z}^{(1)}(\omega, k) \ , \\
\mu^{(1)} (k, \omega) &= \frac{p^{(1)}}{c^2_{s}} \ ,\\
E_{y}^{(1)}(k, \omega) &= - v_{z}^{(1)}(\omega, k)B_{x}^{(0)} + v_{x}^{(1)}(\omega, k)B_{z}^{(0)}\ , \\
j_{y}^{(1)}(k, \omega) &= -\frac{i\omega}{B^{(0)}\cos\theta}\frac{\Gamma p^{(0)}}{c^2_{s}} v_{x}^{(1)}(\omega, k) \ ,
\end{align}
\end{subequations}
since that $\Gamma$ is the adiabatic polytropic index and, remembering when necessary, $u_{f}^2u_{s}^2=(c_{s}^2u^2_{A_{\parallel}})/[(1-u_{A}^2)]$ and $u_{f}^2 + u_{s}^2 = u_{m}^2 + c_{s}^2u_{A_{\parallel}}^2/(1-u_{A}^2)$.

\begin{figure}[ht]
\includegraphics[scale=0.85]{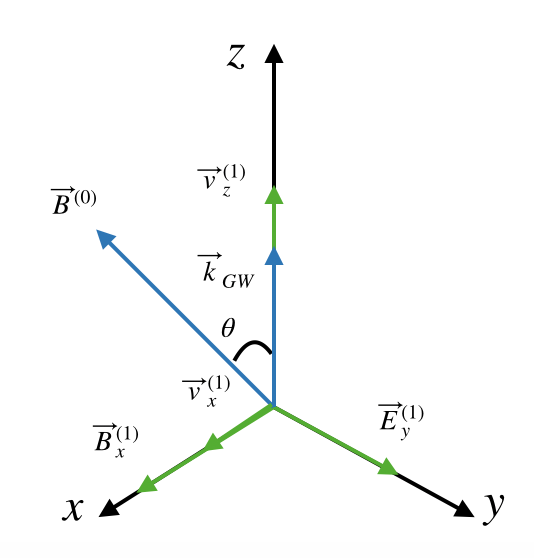}
\caption{\label{fig_xz_plane_MSW} The background magnetic field is on the $x-z$ plane and making a $\theta$ angle with the GW direction. The polarization $h_{+}$ excites the plasma, producing oscillations in the electromagnetic fields ($E^{(1)}_y$ and $B^{(1)}_x$). The oscillations produce compression of the magnetic field lines. This is the magnetoacoustic mode \cite{91}.}
\end{figure}

The directions of the perturbed electromagnetic fields and other physical parameters are shown in Fig. \ref{fig_xz_plane_MSW}. The GW-MSW interaction is more efficient when $\theta\rightarrow\pi/2$, this is because of $v^{(1)}_z$ and $B^{(1)}_x$ being dependent on $\sin{\theta}$. The MSW phase velocity is calculated by Eq.~(\ref{eqnModo:2}). Usually, $c_{s}\ll c$ and the parameter $\sigma\rightarrow0$ in Eq.~(\ref{eqnSigma}).

The slow MSW phase velocity is null [see Eq.~(\ref{eqnModo:2})] and, independent of $\theta$ value, the fast MSW phase velocity produces $u_{f}\rightarrow u_{A}$. If $u_{A}$ in the strongly magnetized plasma reaches the value of $c$, then the fast MSW and the GW can be in coherence, and so it would be possible to transfer a large quantity of energy from the GWs to the plasma.

The constraint on the coherence of the waves is preserved because $\nabla\cdot\vec{B}^{(1)}=0$ and $\vec{k_{f}}\cdot\vec{u_{A}}=0$. Thus, the fast MSW is incompressible and it is able to maintain coherence with the GW (see, e.g. \cite{95}). See that for $\theta=0^{\circ}$, we have $v^{(1)}_z=v^{(1)}_x=0$ and, hence, $B^{(1)}_x=0$.

The charged matter density is absent in the set of perturbed quantities shown in Eqs.~(\ref{eqnGrandezasLaplaceMagnetoCoal}), because it is not perturbed by the propagation of the magnetoacoustic wave. 

\subsection{Rest frame}
\label{subsec_LabFrame}

The equations obtained in the last section are in the comoving frame. However, we must consider the rest energy of the stars when we calculate the energy stored in the plasma by GWs generated through the evolution of the binary system. To do this, we need to apply the Lorentz transformation in order to place the equations in the laboratory frame. Consider that the relativistic wind flows in the $z$-direction (in fact, there is no difference if we choose another direction for the boost).

The Lorentz factor is given by $\gamma = 1/(1-\beta^2)^{1/2}$ while the $\beta$ parameter can be calculated by adding two components. The first component is the velocity of the plasma that is distributed around the NS binary [see Eq.~(\ref{eqnVelOrb})]. The second component is the perturbed velocity $v_{z}^{(1)}$ obtained through interaction with the GWs [see Eq.~(\ref{eqnVelLaplaceMagnetoCoal})]. The plasma is frozen in the magnetic field lines, consequently, the plasma velocity has the same velocity as the stars that move in the binary system.

The perturbed variables are in the frequency domain and so we can proceed in two ways: either the Lorentz transformation is directly applied for the variables in the frequency domain or the parameters of the plasma are converted for the time domain by the Inverse Fourier Transform (IFT) and we apply the Lorentz transformation for returning to the frequency domain. This paper follows the last one; thereby, the interpretation of the variables in the time domain is easier than the first case.

The MHD phase velocities transform like $u^{'}_{A_{\perp}} = u_{A_{\perp}}/[\gamma(1-\beta u_{A_{\parallel}})]$, $u^{'}_{A_{\parallel}}=(u_{A_{\parallel}} - \beta)/(1-\beta u_{A_{\parallel}})$, $u^{'}_{f} = (u_{f}-\beta)/(1-\beta u_{f})$, and $u^{'}_{s} = (u_{s}-\beta)/(1-\beta u_{s})$. From now on, the quantities in the comoving frame are represented with prime ($'$). The boosted velocities associated with the plasma perturbations are given by $v^{'}_{z} \approx \gamma^2v_{z}$ and $v^{'}_{x,y} \approx \gamma v_{x,y}$.  

The Lorentz transformations for the electromagnetic fields are given by $E_x=\gamma(E^{'}_x+\beta B_y^{'})$, $E_y = \gamma(E_y^{'}-\beta B_x^{'})$, $E_z=E_z^{'}$, $B_x = \gamma (B_x^{'}-\beta E_y^{'})$, $B_y=\gamma(B_y^{'}+\beta E_x^{'})$, and $B_z = B_z^{'}$. For an ultrarelativistic plasma, we have $\beta\rightarrow1$. Thus, the Lorentz transformation produces $B_{x}\approx E_{y}$ and $B_{y}\approx E_{x}$. Thereby, the electromagnetic energy integral $\int(|B
^{(1)}_{x}|^2/(2\mu_0)+2|E^{(1)}_y|^2\epsilon_0) d^3k\simeq \int 2 |B_x^{(1)}|^2/(2\mu_0) d^3k$ or $\int(|B
^{(1)}_{y}|^2/(2\mu_0)+2|E^{(1)}_x|^2\epsilon_0) d^3k\simeq \int 2 |B_y^{(1)}|^2/(2\mu_0) d^3k$.

In the rest frame, the background magnetic field is rewritten as $\vec{B}^{(0)} = (\gamma B^{(0)'}_x,0,B^{(0)'}_z)$. Thus, in the laboratory frame, in addition to the poloidal component for the magnetic field, the toroidal component also appears. A part of this behavior can be seen in the simulations developed by \cite{64,96}.

Note that the background electric field does not vanish in the laboratory frame and it has value $\vec{E}^{(0)}=(0,-\beta\gamma B^{(0)'}_x,0)$. The charge and current densities are boosted in the $z$-direction. They are determined by $\rho^{(1)}=\gamma(\rho^{(1)'}+\beta j_z^{(1)'})$, $j^{(1)}_z=\gamma(\beta\rho^{(1)'}+j_z^{(1)'})$, $j^{(1)}_x=j^{(1)'}_x$, and $j^{(1)}_y=j^{(1)'}_y$. Close to the merger and in the ultrarelativistic regime ($\beta\rightarrow1$), we have $\rho^{(1)}(k,\omega) \approx j^{(1)}_{z}(k,\omega)$.

\subsubsection{Relativistic Alfv\'en waves}
\label{subsubsecAlfvenBoostCoal}

For the Alfv\'en mode, the velocity of the perturbation coupled with the GWs in the rest frame is given by

\begin{equation}
\label{eqnVLaplaceBoostCoal}
v_{y}^{(1)}(k, \omega) = -\frac{h_{\times}(\omega_{GW})}{2 \gamma^2}\frac{\omega u_{A_{\perp}}}{(1-\beta u_{A_{\parallel}})}\frac{\omega+ku_{A_{\parallel}}}{\omega^2-k^2u^2_{A_{\parallel}}} \ .
\end{equation}

See that $v_y^{(1)}(k,\omega)$ remains in the same $y$-direction and that it depends on the velocity of the stars and Alfv\'en phase velocity due to the term $(1-\beta^2)/(1-\beta u_{A_{\parallel}}/c)$ (with $c$ explicitly placed in the expression).

For $\theta\rightarrow0$, the perturbed velocity in the rest frame returns to the expression of the comovel frame. However, the perturbed magnetic field decreases, see Eq.~(\ref{eqnBLaplaceBoostCoal}) below, with this coherence condition. In general, for other $\theta$ values, we have $(1-\beta^2)/(1-\beta u_{A_{\parallel}}/c) < 1$.

The component of the perturbed magnetic field stays $\vec{B}^{(1)}=(0,B^{(1)}_y,0)$ and $B_{y}^{(1)}$ is given by

\begin{align}
\label{eqnBLaplaceBoostCoal}
B_{y}^{(1)} (k, \omega) = \frac{h_{\times}(\omega_{GW})B^{(0)}_{x}}{2 (1-\beta u_{A_{\parallel}})} \frac{\omega^2+k^2 u_{A_{\parallel}}^2}{\omega^2-k^2 u_{A_{\parallel}}^2} \ .
\end{align}

The components of the perturbed electric field stay $\vec{E}^{(1)}=(E^{(1)}_x,0,E^{(1)}_z)$ where the expressions for $\vec{E}^{(1)}_{x,z}$ are 

\begin{subequations}
\label{eqnELaplaceBoostCoal}
\begin{align}
    E_{z}^{(1)} (k, \omega)&= \gamma^2 B_{x}^{(0)}v_{y}^{(1)}(k, \omega) \ , \\ 
    E_{x}^{(1)} (k, \omega) &= \gamma[\beta B^{(1)'}_y - \gamma B_{z}^{(0)}v_{y}^{(1)}(k, \omega)] \label{eqnELaplaceBoostCoal:Ex}\ .
\end{align}
\end{subequations}

Note that during the binary coalescence, the components of the electric field increase, on turn, the current densities in these directions also increase, mainly, the perturbed electric field and current density associated to the $x$-components. The charge $\rho^{(1)}$ and current densities $j^{(1)}_z$ and $j^{(1)}_x$ are given, respectively, by

\begin{subequations}
\label{eqndensitiesAW}
\begin{align}
    \rho^{(1)}(k,\omega)&=\frac{i\gamma^2}{4\pi}\frac{E^{(1)}_z}{u_{A_{\parallel}}-\beta}\bigg[u_{A_{\parallel}}(k+\beta\omega)-\beta(ku_{A_{\parallel}}^2+\beta\omega)\bigg] \\
    j^{(1)}_z(k,\omega)&=\frac{i\gamma^2}{4\pi}\frac{E^{(1)}_z}{u_{A_{\parallel}}-\beta}\bigg[u_{A_{\parallel}}(\beta k+\omega)-\beta(\beta ku_{A_{\parallel}}^2+\omega)\bigg] \\
    j^{(1)}_x(k,\omega) &= \frac{-i\omega\gamma^2}{4\pi}\bigg[\frac{(1-\beta u_{A_{\parallel}})^2}{(u_{A_{\parallel}}-\beta)^2}-1\bigg]\nonumber \\
    &\times\big[E^{(1)}_x(k,\omega)-\beta B^{(1)}_y(k,\omega)\big] \ . 
\end{align}
\end{subequations}

The directions of the perturbed electromagnetic field components remain the same as shown in Fig. \ref{fig:xz_plane_AW}. The $y$-component of the background electric field appears due to the boost. It was added to the term $B^{(1)'}_y$ in the $x$-component of the perturbed electric field according to Eq. ~(\ref{eqnELaplaceBoostCoal:Ex}). For $\beta\rightarrow 0$, Eqs.~(\ref{eqnBLaplaceBoostCoal}), (\ref{eqnELaplaceBoostCoal}), and (\ref{eqndensitiesAW}) return to those written in the proper frame.

Note that the perturbed magnetic field has maximum value when $\theta=\pi/2$ and the other plasma parameters also increases with $\sin(\theta)$. On the other hand, the Alfv\'en phase velocity increases with $\cos(\theta)$. Close to the merger, the plasma is ultrarelativistic, i.e., $u_{A}\rightarrow 1$ ($c=1$). This coherence condition is established if $\theta\rightarrow0$.

However, in this condition, the perturbed plasma parameters become small and this causes the associated energy to reach low values, even with the increase in the frequency and amplitude of the GWs.

The Alfv\'en wave mode is not the more efficient MHD mode to interact with the GWs, considering that GW phase velocity remains near to the speed of light when crossing the region where the plasma is distributed \cite{78}. 

\subsubsection{Relativistic magnetosonic waves}

For the magnetoacoustic mode, the perturbed velocities coupled to the GWs are

\begin{widetext}
\begin{subequations}
\label{eqnVLaplaceBoostCoal_2}
\begin{align}
v_{z}^{(1)}(k,\omega) &= \frac{h_{+}}{2\gamma^2}\frac{\omega u_{A_{\perp}}^2}{(1-\beta u_{A_{\parallel}})^2} \frac{(1-\beta u_{f})(1-\beta u_{s})}{(u_{f}-u_{s})\big[(u_{f}+u_{s})(1+\beta^2)-2\beta(1+u_{f}u_{s})\big]} \cdot \Bigg\{u_{f}(1-\beta u_{s})\bigg[\frac{u_{f}k(1-\beta u_{f})+\omega (u_{f}-\beta)}{\omega^2-k^2 u_{f}^2}\bigg] \nonumber \\  -&u_{s}(1-\beta u_{f})\bigg[\frac{u_{s}k(1-\beta u_{s})+\omega(u_{s}-\beta)}{\omega^2-k^2u_{s}^2}\bigg]\Bigg\} \ , \\
v_{x}^{(1)}(k,\omega) &= -\frac{h_{+}}{2}\frac{\omega u_{A_{\perp}}(u_{A_{\parallel}}-\beta)c_{s}^2}{1-u_{A}^2}\frac{(1-\beta u_{f})^2(1-\beta u_{s})^2}{(u_{f}-u_{s})\big[(u_{f}+u_{s})(1+\beta^2)-2\beta(1+u_{f}u_{s})\big]} \cdot \Bigg\{\frac{1-\beta u_{s}}{(u_{s}-\beta)^2}u_{s}\frac{[u_{s}k(1-\beta u_{s})+\omega(u_{s}-\beta)]}{\omega^2-k^2u_{s}^2} \nonumber \\  -& \frac{1-\beta u_{f}}{(u_{f}-\beta)^2}u_{f}\frac{[u_{f}k(1-\beta u_{f})+\omega(u_{f}-\beta)]}{\omega^2-k^2u_{f}^2}\Bigg\} - \gamma^2 \bigg(\frac{u_{A_{\parallel}}-\beta}{u_{A_{\perp}}}\bigg)v_{z}^{(1)}(k,\omega) \ ,
\end{align}
\end{subequations}
\end{widetext}
where $\tan{\theta}'={u_{A_{\perp}}}'/{u_{A_{\parallel}}}'=u_{A_{\perp}}/[\gamma(u_{A_{\parallel}}-\beta)]$ and $1-{u_{A}}^{2'}=(1-u_{A}^2)/[\gamma^2(1-\beta u_{A_{\parallel}})^2]$.

At this point, we need to be more careful because the magnetic $B_{x}^{(1)}$ field component is the biggest and several steps are needed to obtain the final equation. First, the boosted expression is $B_{x} =\gamma\big[{B_{x}}'-\beta {E_{y}}'\big]$ and the right side terms are given by

\begin{widetext}
\begin{subequations}
\begin{equation}
E_{y}^{(1)}(k, \omega) = - v_{z}^{(1)}(\omega, k)B_{x}^{(0)} + v_{x}^{(1)}(\omega, k)B_{z}^{(0)} \label{eqnECartCoalCoMovel} \ ,
\end{equation}
\begin{equation}
\frac{B^{(1)}_{x}(k, \omega)}{B^{(0)}} = v_{z}^{(1)}(\omega,k)\sin\theta - v_{x}^{(1)}(\omega, k)\cos\theta - \frac{\omega}{\omega+k}\frac{1-u^2_{A}}{u^2_{A}}\frac{v^{(1)}_{x}(\omega, k)}{\cos\theta}\label{eqnBCartCoalCoMovel} \ ,
\end{equation}
\end{subequations}
\end{widetext}
where the prime ($'$) is hidden to avoid mess. The expression for $E_{y}^{(1)}$ results in

\begin{equation}
\label{eqnELaplaceCoal}
E_{y}^{(1)}(k,\omega)= -\bigg[B_{x}^{(1)}(k,\omega)+\frac{\omega}{\omega+k}\frac{1-u_{A}^2}{u_{A}^2}\frac{B^{(0)}}{\cos\theta}v_{x}^{(1)}\bigg] \ .
\end{equation}

Using the Laplace transformation in the $z-t$ space, we have

\begin{align}
\label{eqnECartCoal}
E_{y}^{(1)}(z,t)=& -\Bigg[B_{x}^{(1)}(z,t)\nonumber \\+&\frac{1-u_{A}^2}{u_{A}^2}\frac{B^{(0)}}{\cos\theta}\mathcal{L}^{-1}\bigg\{\frac{\omega}{\omega+k}v_{x}^{(1)}(k, \omega)\bigg\}\Bigg] \ .
\end{align}

Defining ${\hat{v}_{x}^{(1)}}(z,t)=\mathcal{L}^{-1}\bigg\{\omega/(\omega+k)v_{x}^{(1)}(k, \omega)\bigg\}$ and returning the prime ($'$), we obtain

\begin{equation}
\label{eqnECartBoostCoal}
E_{y}^{(1)\,'}(z,t)= -\bigg[{B_{x}^{(1)'}}(z,t)+\frac{1-{u_{A}^{2\,'}}}{{u_{A}^2}'}\frac{{B^{(0)\,'}}}{\cos{\theta}'}{\hat{v}_{x}^{(1)'}}(z,t)\bigg] \ .
\end{equation}

Now, returning to the expression $B_{x} =\gamma\big[{B_{x}}'-\beta {E_{y}}'\big]$ and using Eq.~(\ref{eqnECartBoostCoal}), we find

\begin{equation}
\label{eqnBCartBoostCoal}
B_{x}^{(1)}(z,t) = \gamma \bigg[{B_{x}^{(1)}}'(z,t)(1+\beta)+\frac{\beta}{\gamma}\frac{1-u^2_{A}}{(u_{A_{\parallel}}-\beta)^2} B_{z}^{(0)}{\hat{v}_{x}^{(1)\,'}}(z,t)\bigg] \ .
\end{equation}

Finally, using Eqs.~(\ref{eqnBCartCoalCoMovel}) and~(\ref{eqnECartBoostCoal}) in the Fourier space, we obtain the perturbed magnetic field for the MSW as

\begin{widetext}
\begin{equation}
\label{eqnBLaplaceBoostCoal_2}
B_{x}^{(1)}(k,\omega) = -\frac{h_{+}}{2\gamma^2}\frac{\omega(1-\beta u_{f})^2(1-\beta u_{s})^2}{(u_{f}-u_{s})\big[(u_{f}+u_{s})(1+\beta^2)-2\beta(1+u_{f}u_{s})\big]} \big[\Xi(k,\omega)+\Pi(k,\omega)\big] \ .
\end{equation}
\end{widetext}

The expressions for $\Xi(k,\omega)$ and $\Pi(k,\omega)$ are in the \ref{apen}. The term $\Xi(k, \omega)$ becomes relevant when the GW propagates perpendicular to the background magnetic field ($\theta =\pi/2$). For this condition, the term $\Pi(k,\omega)$ becomes less relevant. The opposite occurs when $\theta=0$. 

The components of the perturbed electric field stay $\vec{E}^{(1)}=(0,E^{(1)}_y,0)$, where the expression for ${E}^{(1)}_{y}$ is

\begin{equation}
\label{eqnELaplaceBoostCoalMSW}
    E_{y}^{(1)} (k, \omega) = \gamma[E_{y}^{(1)'}(k, \omega)-\beta B^{(1)'}_x(k, \omega)] \ .
\end{equation}

The directions of the perturbed electromagnetic field components remain the same as shown in Fig.~\ref{fig_xz_plane_MSW}. We only added the term $B^{(1)'}_x$ in the $y$-component of the perturbed electric field. 

The mechanical pressure and the current density in the rest frame are, respectively, given by

\begin{subequations}
\label{eqnjpboostmsw}
\begin{align}
    p^{(1)} (k,\omega)&=\frac{ku_{f}}{\omega}\frac{\gamma(1-\beta u_f)}{u_f-\beta}\Gamma p^{(0)}\gamma^2v_z^{(1)}(k,\omega) \\
    j^{(1)}_y(k,\omega)&=\frac{-i\omega \gamma^2}{B^{(0)}_z}\frac{\Gamma p^{(0)}}{c_s^2}v^{(1)}_x(k,\omega)
\end{align}
\end{subequations}

For $\beta\rightarrow 0$, Eqs.~(\ref{eqnBLaplaceBoostCoal_2}), (\ref{eqnELaplaceBoostCoalMSW}), and (\ref{eqnjpboostmsw}) return to the proper frame.

According to Eq.~(\ref{eqnModo:2}), with $c_{s}\ll c$, the fast MSW phase velocity $u_{f}\rightarrow c$ and the slow MSW phase velocity $u_{s}\rightarrow 0$. These results are independent of $\theta$. Therefore, the fast MSW mode is more efficient than the slow MSW mode for interacting coherently with the GWs. While the binary stars coalesce, the perturbed magnetic field $B^{(1)}_x$ increases with the resonant interaction, and so the toroidal magnetic field component also increases.

The electric field produces a current density in the $y$-direction. In the last topic above, we show that $j_x^{(1)}$ increases with the inspiral phase. Thereby, in the $x-y$ plane, arises a current density responsible for producing a magnetic field component in the $z$-direction. This new magnetic component is parallel to the poloidal background magnetic field of the compact binaries. This behavior is also found in \cite{64,96}. 

From now on, we return to the international system of units (SI) with the necessary changes in the MHD equations. The main replacement rules for translating the equations from Gaussian to SI are: $\vec{E}\rightarrow\sqrt{4\pi\epsilon_0}\, \vec{E}$, ($\rho,\vec{j})\rightarrow1/\sqrt{4\pi\epsilon_0}\,(\rho,\vec{j}$), and $\vec{B}\rightarrow\sqrt{4\pi/\mu_0}\,\vec{B}$.

\section{Stored Energy by the Coupling}
\label{secEnergy}

In this section we will assess how the energy is transferred from GWs to EMWs during the inspiral phase down to the merger. The NSs produce a wind composed by electron-positron pairs and also a plasma satisfying the ideal MHD theory. This plasma is filling in the space up to large distances.

We estimate the magnitude of the MHD energy modes excited by the amplitudes of the GWs and including the evolution of the perturbed magnetic field and of the plasma velocities. The stars have an initial magnetic field, in the comoving frame, that is anchored in the polar caps of each star. It evolves as a dipole, that is, $B(r) = B_{\star}(R_{\star}/r)^3$, where $B_{\star}$ and $R_{\star}$ are parameters on the surface of the stars \cite{97,98,99}.

The NSs orbit each other with their magnetic fields. The maximum of the magnetic field of the system remains nearly constant up to $\sim 3\,{\rm ms}$ before the merger (see, e.g., \cite{64}). However, it decreases with the distance as a magnetic dipole \cite{96}.

We consider that the magnetic dipole moments ($\mu=B_{\star}R_{\star}^3$) are aligned with the orbital angular momentum during the evolution of the system. In \cite{59}, the authors consider three different configurations of the magnetic moments with respect to the orbital angular momentum. The configurations are parallel, antiparallel with the same moments $\mu$, and parallel with different magnetic moments (in this last case, $\mu_1=100\,\mu_2$). The authors find that the antiparallel case is significantly more radiative than the other cases, especially in the late stage of the coalescence.  

The accelerated motion of the NSs induces a wind that is corotating with the stars. In the MHD approximation, the plasma is frozen in the magnetic field lines and is forced to high velocities. The radius at which the tangential linear velocity is equal to $c$ is the light cylinder radius given by $R_{LC} = c/\Omega$ \cite{99,100}, being $\Omega$ the orbital frequency of the system. The magnetosphere of the BNSs extends up to $R_{LC}$, and for greater distances the wind is free of forces. 

The morphology of the wind is determined by the magnetic field geometry, since the charged particles are frozen in the magnetic field lines. The charged matter density is calculated by the Goldreich-Julian density \cite{80}:

\begin{equation}
\label{eqnDensidadeGJ}
n_{GJ} = \frac{2 \epsilon_0}{\left |  e \right |} \vec{\Omega}\cdot\vec{B} \ ,
\end{equation}
where $\epsilon_0$ is the vacuum permittivity, $e$ is the elementary charge, and $B$ corresponds to the background magnetic field.

For the canonical binary system close to the merger, we have $\Omega=f_{GW}/2=750$ Hz. Considering $B=10^{8}\,{\rm T}$, we have $n_{GJ}\sim 10^{19}\,{\rm m}^{-3}$. Far from the stars, the magnetic field lines start to open and the particles flow out, unbalancing the Goldreich-Julian density. As a result, a strong electric field appears to the longitudinal direction and the ``primary'' charged particles can flow out from the star surface with high Lorentz factor ($\gamma_{p} \sim 10^7$).

Due to the geometry of the magnetic field lines and to the inverse Compton scattering \cite{101,102}, a cascade effect is created. As a consequence, ``secondary'' $e^{\pm}$ pairs are produced with typical density $n_{s}=M n_{p}$, where $M$ is the so-called ``multiplicity factor''. By the energy conservation, the Lorentz factor of the ``second'' particle generation is $\gamma_{s}=\gamma_{p}/M$ and it can reach values $\gamma_{s} \sim 10^2$ \cite{103}.

Based on the above discussion, the morphology of the BNS magnetosphere is determined by the dipolar magnetic field and it extends up to the light cylinder radius. Thus, the GWs generated by the stars during coalescence excite the plasma. As a consequence, electric currents are generated.

The plasma perturbation produces the MHD modes and the GW-EMW interaction in the vacuum, or in a medium, is more efficient when the phase velocities of the waves (GWs and EMWs) are in coherence. We calculate the electromagnetic energy excited by the interaction using Parseval's Theorem. The electromagnetic field stores a quantity of energy which can be calculated as \cite{92}

\begin{equation}
\label{eqnBFourier}
W^{(B)}=\int_{\Re_{\vec{r}}} \frac{\left | B(\vec{r}) \right |^2}{2\mu_{0}} d^3r = \int_{\Re_{\vec{k}}} \frac{\left | B(\vec{k}) \right |^2}{2\mu_{0}}d^3k \ ,
\end{equation}
where $\mu_{0}$ is the vacuum permeability and the magnetic field spectral density is evaluated over the entire frequency region $\mathfrak{R}_{\vec{k}}$.

For the Alfv\'en mode, the perturbed magnetic field in the frequency domain, see Eq.~(\ref{eqnBLaplaceBoostCoal}), and in units of SI is

\begin{equation}
\label{eqnBySimplified}
B_{y}^{(1)} (k, \omega) = \frac{h_{\times}(\omega_{GW})B^{(0)}_{x} c}{2 (1-\beta u_{A_{\parallel}}/c)} \frac{\omega^2+k^2 u_{A_{\parallel}}^2}{\omega^2-k^2 u_{A_{\parallel}}^2} \ .
\end{equation}

Thus, the integral in Eq.~(\ref{eqnBFourier}) can be calculated by considering a wave number sphere of radius in the range $[0,k]$. The volume element in the frequency domain is $d^3k=4\pi k^2 dk$ and we have for the electromagnetic energy of the Alfv\'en wave the following result:

\begin{widetext}
\begin{align}
\label{eqnWBAW}
W_{AW}^{(B)} = \Bigg|\frac{B_{x}^{(0)^2} h_{\times}^2\pi c^2}{2\mu_{0}(1-\beta u_{A_{\parallel}}/c)^2}\Bigg(\frac{\omega_{AW}}{u_{A_{\parallel}}}\Bigg)^3\times\Bigg\{\frac{1}{3}+\frac{2\omega^2}{\omega_{AW}^2}\Bigg(2+\frac{\omega^2}{\omega^2-\omega^2_{AW}}\Bigg)- \frac{6\omega^3}{\omega_{AW}}\tanh^{-1}\Big(\frac{\omega_{AW}}{\omega}\Big)\Bigg\}\Bigg| \ .
\end{align}
\end{widetext}

Observe that the electromagnetic energy of the GW-EMW interaction depends on the following physical parameters: the GW amplitudes, the GW-EMW frequencies, the background magnetic field vector, the Alfv\'en phase velocity, and the $\beta$ factor in the rest frame. See that $\beta$ is the sum of the velocity of the stars with the velocity of the perturbed plasma. 

For the magnetosonic mode, we need to simplify the expression of the perturbed magnetic field [see Eq.~(\ref{eqnBLaplaceBoostCoal_2})]. Consider the phase velocities $u_{f,s}$ [see Eq.~(\ref{eqnVelocidadeFase})] in the ultrarelativistic limit, where $u_{A}\rightarrow c$. As discussed previously, the phase velocity of the slow magnetosonic mode is $u_s\rightarrow0$. On the other hand, the fast magnetosonic mode can be in coherence with the GWs, where $u_f^2\approx u^2_{A}$, when $c_{s} \ll c$. 

Taking $u_{A_{\parallel}}=u_{A}\cos(\theta)$ and $u_{A_{\perp}}=u_{A}\sin(\theta)$, we obtain to the perturbed magnetic field in the frequency domain

\begin{widetext}
\begin{align}
\label{eqnBxSimplified}
   B^{(1)}_x \simeq &\frac{h_{+}B^{\text{(0)}}}{2 c(1-\beta) \gamma ^4}\frac{\omega}{\left(k^2 u_A^2-\omega
   ^2\right)}\frac{\left(1-\frac{\beta  u_A}{c}\right)}{\left(\beta ^2 u_A+u_A-2 \beta  c\right) \left[\beta  c-u_A \cos (\theta)\right]^2 \left[c-\beta  u_A \cos
   (\theta)\right]^2} \nonumber \\ &
   \times \Bigg\{(1-\beta ) \beta  \gamma  \omega  \cos (\theta) \left(c^2-u_A^2\right)
   \left(u_A-\beta  c\right) \left[u_A-u_A \cos (\theta)\right] \bigg[\left(\beta
   ^2+1\right) c u_A [\cos (\theta)+1]-2 \beta  \left[u_A^2 \cos
   (\theta)+c^2\right]\bigg] \nonumber \\ &-\sin (\theta) \bigg[2 \beta ^2 c^3 \Big\{\gamma ^2 k u_A
   \left(c-\beta  u_A\right) \left[\beta  c-u_A \cos (\theta)\right]^2+\omega 
   \left(u_A-\beta  c\right) \left[\gamma ^2 \left[c-\beta  u_A \cos
   (\theta)\right]^2-u_A^2 \sin ^2(\theta)\right]\Big\} \nonumber \\ &+c \left[\beta  c-u_A \cos(\theta)\right]^2 \Big\{k u_A \left(c-\beta  u_A\right) \left[\gamma ^2 \left[\beta c-u_A \cos (\theta)\right]^2+u_A^2 \sin ^2(\theta)\right]+\gamma ^2 \omega \left(u_A-\beta  c\right) \left[c-\beta  u_A \cos
   (\theta)\right]^2\Big\}\bigg]\Bigg\} \ .
\end{align}
\end{widetext}

By doing $\theta = \pi/2$ in Eq. (\ref{eqnBxSimplified}), we retrieve the equation for the magnetosonic mode seen earlier in which the propagation of the GWs is perpendicular to the background magnetic field. So, for $\theta = \pi/2$ we obtain

\begin{widetext}
\begin{align}
   B^{(1)}_x(k,\omega)=&\frac{h_{+}B^{\text{(0)}}}{2 (\beta -1) c^2 \gamma ^4} \frac{\omega}{\left(\omega^2-k^2 u_A^2 \right)}\frac{\left(1-\beta  u_A/c\right)}{\left(\beta ^2 u_A+u_A-2
   \beta  c\right)} \nonumber \\ &\times\bigg\{3 c^2 \gamma ^2
   \left[-u_A \left(\beta ^2 c k+\omega \right)+\beta ^3 k u_A^2+\beta  c \omega
   \right]+u_A^2 \left[u_A \left(\beta  k u_A+2 \omega \right)-c \left(k u_A+2 \beta 
   \omega \right)\right]\bigg\} \ .
\end{align}
\end{widetext}

Integrating Eq.~(\ref{eqnBxSimplified}) over all frequency space $k^2dk$, we obtain the electromagnetic energy excited by GW-fast MSW interaction. This result is

\begin{widetext}
\begin{align}
\label{eqnWBMSW}
    W^{(B)}_{MSW}=&\left|\frac{\pi h_{+}^2 B^{\text{(0)}^2}_x}{32\mu_0 u_{A}}\frac{ \omega ^2}{(\omega^2-\omega_{MSW}^2)} \frac{(1-\beta)^2 (\beta+1)^4 \left[4 \cos (\theta )-\beta^2\cos (2 \theta
   )+\beta^2-8\right]^2}{\left(1-u_{A}^2
   \cos (\theta )/c\right)^4} \right| \nonumber \\ &\times\left|\bigg\{4 \omega_{MSW}^3-6 \omega_{MSW} \omega ^2+\omega  (\omega^2 -\omega_{MSW}^2) \left[-7 \log (\omega_{MSW}-\omega )+7 \log (\omega_{MSW}+\omega )-8 \tanh
   ^{-1}\left(\frac{\omega_{MSW}}{\omega }\right)\right]\bigg\}\right| \ ,
\end{align}
\end{widetext}
where we used $\gamma=1/(1-\beta^2)^{1/2}$ and $\omega_{MSW}=k u_{f}$.

For energy functions on real domains, we verify that the terms $1/(\omega^2_{(AW,MSW)}-\omega^2)$, $\tanh^{-1}(\omega_{(AW,MSW)}/\omega)$, and $\log(\omega^2_{(AW,MSW)}-\omega^2)$ go to infinity when the GWs and the AWs have the same frequency $\omega_{(AW,MSW)}-\omega\rightarrow0$.

The wave interactions are most efficient when they are in coherence (resonant) \cite{73,74,78}. Therefore, we will do the analysis taking into account the condition $\omega_{(AW,MSW)} - \omega=\Delta \omega \rightarrow 0$ during the inspiral phase of the system. 

Concerning to the $\theta$ angle, if $\theta\rightarrow0$, then both the perturbed magnetic field and the perturbed electromagnetic energy vanish. This result is expected because when the background magnetic field is parallel to the GW propagation, there is no coupling between GWs and EMWs. In this case, the GW propagates in the same direction of $\vec{B^{(0)}}$ and so no electrical current is produced in the plasma \cite{73,74,78,95}.

In order to apply our formalism, we consider two cases for calculating the electromagnetic energy: the simulations developed by \cite{64} and the event GW170817 \cite{104} where, with our results, a comparison can be made with its electromagnetic counterpart, that is, the associated GRB170817A \cite{105}.

Table~\ref{tableDatas} shows a short summary on the initial conditions of the simulations developed by \cite{64} and the physical parameters inferred from the detection of GW170817. It has been also included in the Table the initial conditions that we apply to the equations of our formalism developed in Sec. \ref{secEquations}.

We consider that the magnetic field on the surface of the stars associated with GW170817 was $10^8\,{\rm T}$. This is considered a realistic value for NSs \cite{59,96,98}. For the equatorial radius of the NSs, we use $1.36\times10^4\,{\rm m}$ \cite{64,98}.

\begin{table}[ht]
\caption{\label{tableDatas} The initial conditions based on the simulations developed by \cite{64} and the source GW170817 \cite{104}. The values for the parameters $R_{0}$, $\tau_0$, $f_{GW_{\rm{ISCO}}}$ and $f_{GW}(\tau=2\rm{ms})$ are the initial conditions obtained from our model.}
\begin{ruledtabular}
\begin{tabular}{lcr}
\textbf{Physical parameter}         & \textbf{Simulation} & \textbf{GW170817} \\ \hline
Total Mass   & $3.0M_{\odot}$                & $2.73 M_{\odot}$           \\ 
Chirp Mass        & $1.31M_{\odot}$             & $1.186 M_{\odot}$           \\ 
$B_{\star}$   & $10^8$ T               & $10^8$ T             \\ 
Individual Radius & $1.36\times10^4$ m          & $1.36\times10^4$ m          \\  
$R_{0}$ & $1.592\times10^5$ m           & $1.543\times10^5$ m          \\
$\tau_0$ & $1.926$ s           & $2.259$ s          \\
$f_{GW_{\rm{ISCO}}}$ & $1.465$ kHz          & $1.610$ kHz         \\ 
$f_{GW}(\tau=2\rm{ms})$ & $1.314$ kHz          & $1.395$ kHz          \\ 
\end{tabular}
\end{ruledtabular}
\end{table}

Our strategy consists, firstly, of using the results of the simulations developed by \cite{64} to ``calibrate'' the discrete inverse Fourier transform to be applied for our GRMHD equations. This allows us to obtain parameters such as, for example, the Alfv\'en velocity value that better reproduces the characteristics of these simulations. This procedure will allow, in the next step, to apply our formalism to the source GW170817. In the case of the Alfv\'en velocity, the comparison with \cite{64} shows that $u_{A}\simeq0.2\,c$. 

It is important at this point to highlight some aspects of these simulations. The initial conditions used by \cite{64} correspond to an initial separation of $4.5\times 10^{4}\,{\rm m}$ between the stars (distance measured between the centers of the two stars). The stars quickly lose angular momentum through the emission of GWs and after $8\,{\rm ms}$ of evolution a hypermassive NS (HMNS) is formed. After $\sim {\rm ms}$, the HMNS loses angular momentum by collapsing to form a black hole with mass of $\sim 2.9\,M_ {\odot}$.

\begin{figure}[ht]
\includegraphics[scale=0.42]{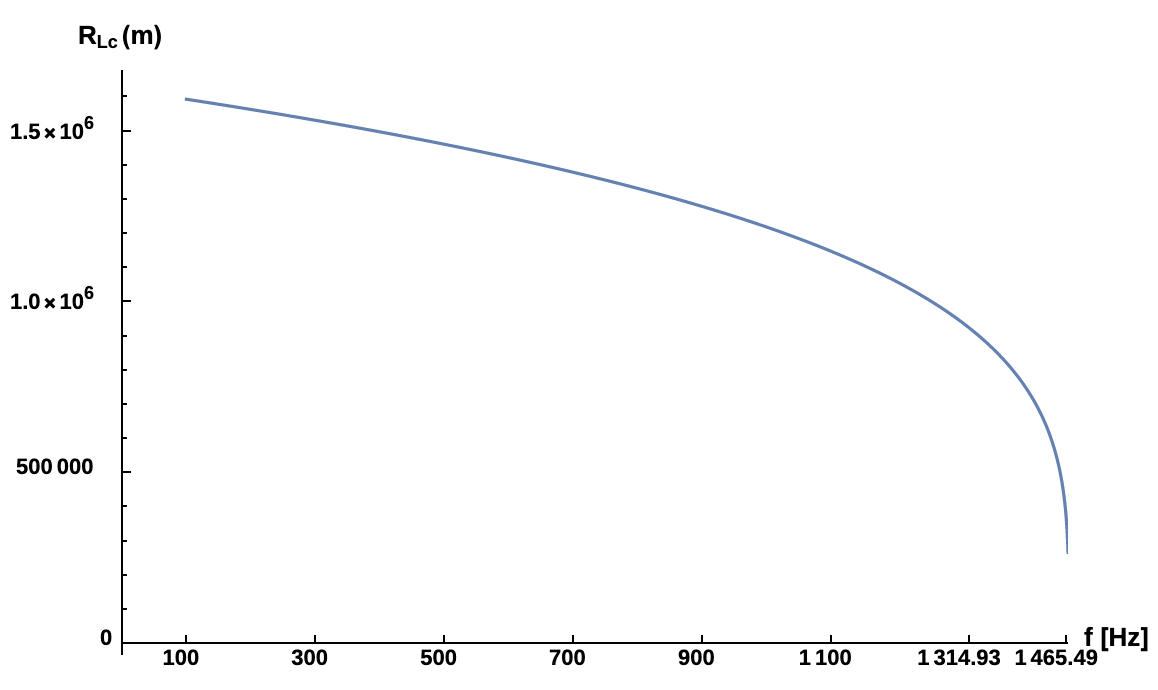}
\caption{\label{fig:RxF} The evolution of the light cylinder radius as a function of $f_{GW}$. We consider the inspiral phase from $100\,{\rm Hz}$ to $1465.49\,{\rm Hz}$ (BNSs at ISCO frequency).}
\end{figure}

Regarding our model, we take the frequency of $100\,{\rm Hz}$ as the initial one. This allows us to obtain $\tau_{0} = 1.926\,{\rm s}$ and $R_{0} = 1.592\times 10^{5}\,{\rm m}$ as indicated in Table \ref{tableDatas}. When $\tau = 2\,{\rm ms}$ the GW frequency reaches the value $1.314\,{\rm kHz}$ while at ISCO the frequency reaches $1.465\,{\rm kHz}$. The emission of GWs causes the orbit to shrink until the NSs make contact, consequently the light cylinder radius decreases (see Fig.~\ref{fig:RxF}).

In the limit of quasicircular orbit, we consider the inspiral phase until to reach the ISCO at $\tau=1.4\,{\rm ms}$ before the merger. The simulations in \cite{64} show that the MHD energy and the toroidal magnetic field have significant values when $\tau\leq 2\,{\rm ms}$ to the merger (see Fig.~2 of these authors).

\begin{figure}[ht]
\includegraphics[scale=0.42]{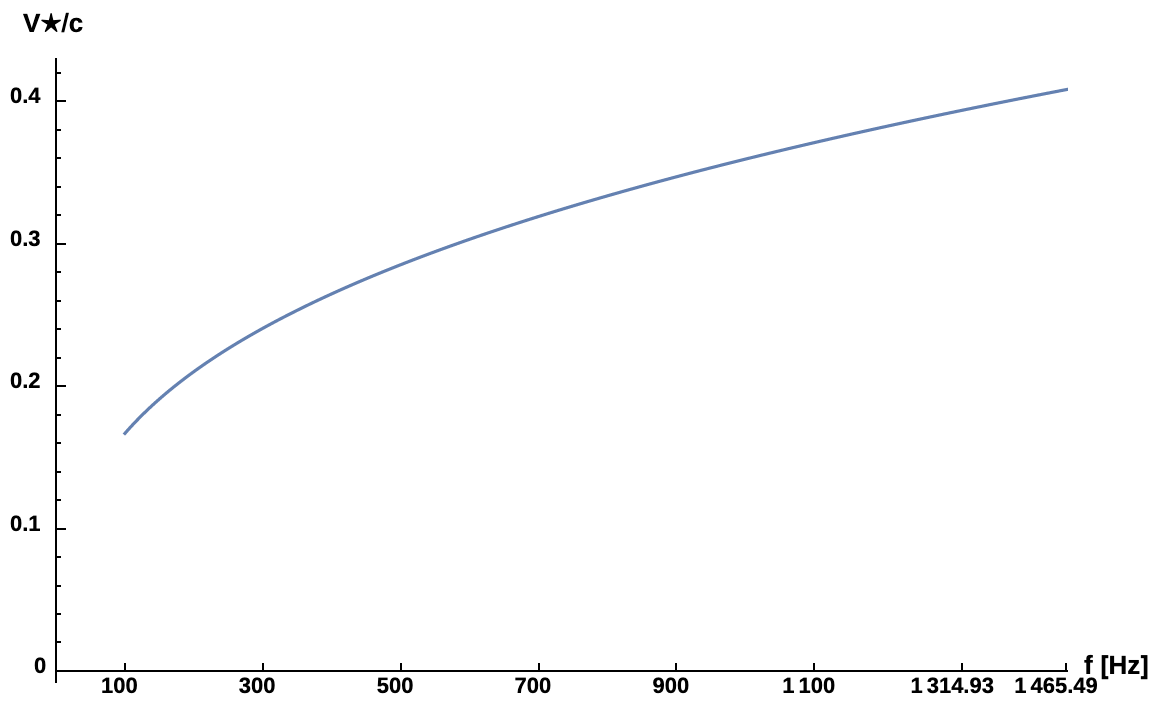}
\caption{\label{fig:vstarsxF} The evolution of $v_{\star}/c$ as a function of $f_{GW}$. It describes a part of the $\beta$ parameter in the Lorentz transformation. The other part is the perturbed plasma velocity that comes from the interaction GW-EMW.}
\end{figure}

The stars reach high velocities before the merger. The value of $v_{\star}$ can be obtained from Eq.~(\ref{eqnVelOrb}) and its relation with the GW frequency is shown in Fig.~\ref{fig:vstarsxF}. Note that $v_{\star}$ is part of the $\beta$ parameter in the Lorentz factor. We need to add the perturbed plasma velocity by GW-EMW interaction with $v_{\star}$ in order to obtain $\beta$.

\begin{figure}[ht]
\includegraphics[scale=0.42]{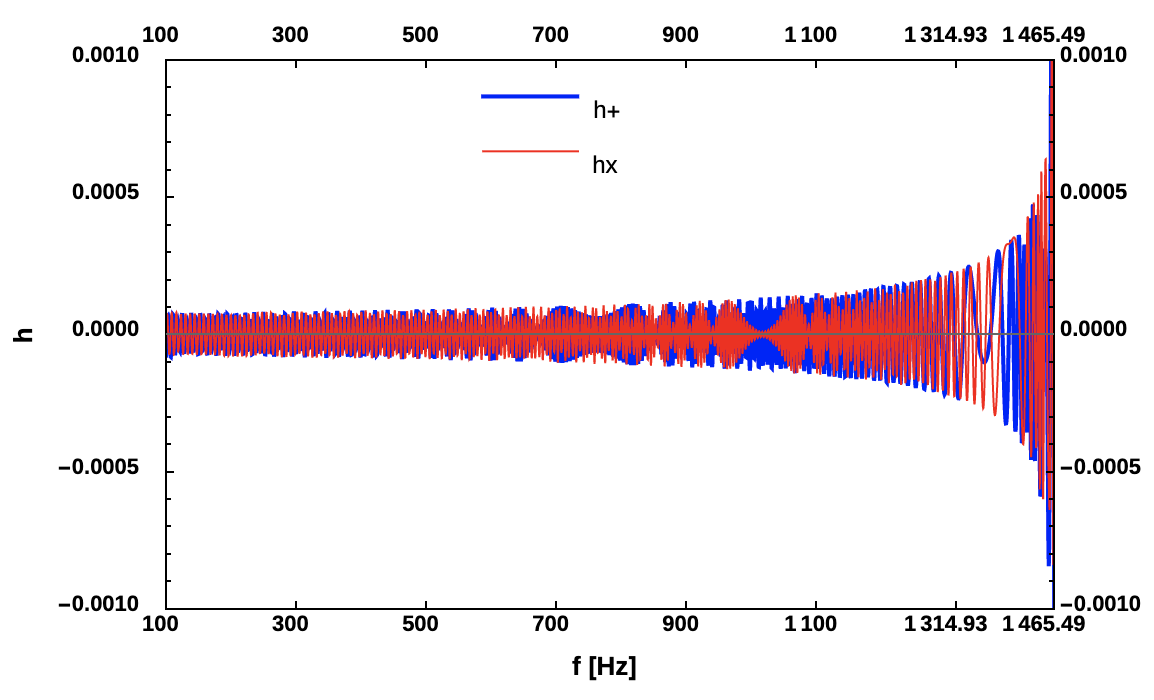}
\caption{\label{fig:hTimexF} The evolution of the GW amplitudes ($h_{+}$ and $h_{\times}$) as functions of $f_{GW}$ and in the time domain. The waveforms are calculated on discrete form (see \cite{88}). The red thin line represents the $\times$-polarization while the blue thick line represents the $+$-polarization.}
\end{figure}

\begin{figure}[ht]
\includegraphics[scale=0.42]{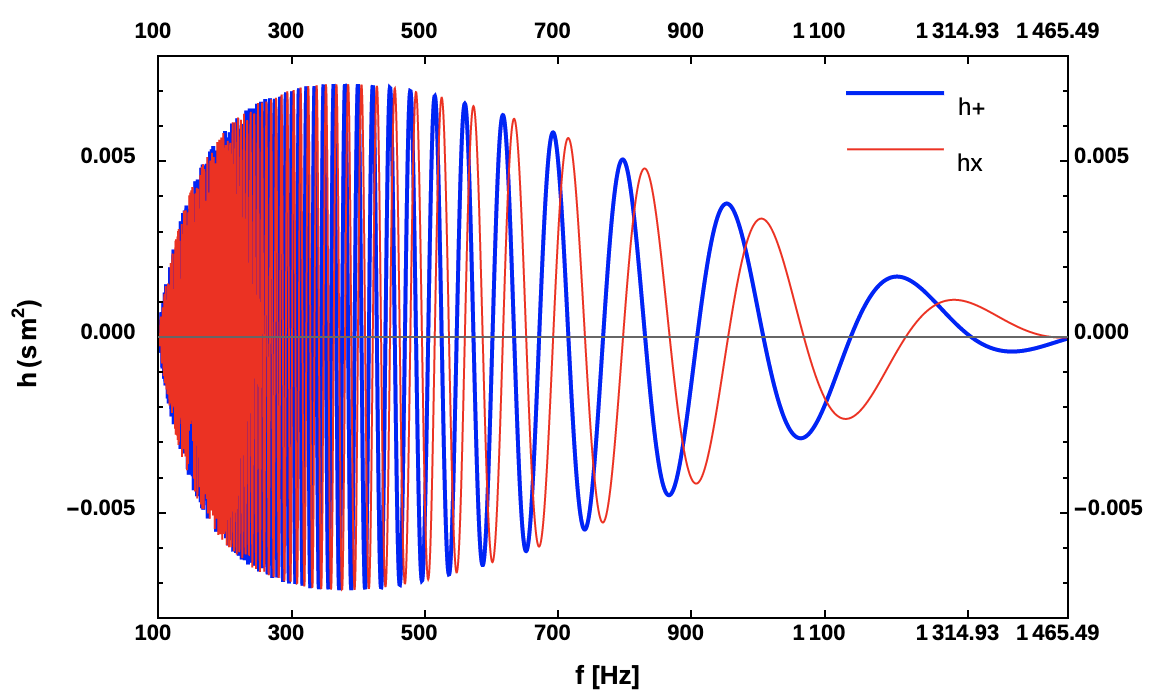}
\caption{\label{fig:hxF} The evolution of the GW amplitudes as functions of $f_{GW}$ and in the frequency domain. The red thin line represents the $\times$-polarization while the blue thick line represents the $+$-polarization.}
\end{figure}

The amplitude and frequency of the GWs increase as the inspiral phase evolves. In Fig.~\ref{fig:hTimexF}, we present the $h_{+}$ and $h_{\times}$ polarizations in the time domain and as functions of the frequency $f_{GW}$. In Fig.~\ref{fig:hxF}, we show how the GW amplitudes evolve in the frequency domain. The region of interaction for the characterization of the GWs is described by the light cylinder of the system. The ordinate axis in Fig.~\ref{fig:hxF} represents the amplitude density or a histogram of the amplitude over the associated frequencies.

We calculate the physical quantities in the frequency domain as the equations are more easily worked out in that domain. Then, we apply the discrete inverse Fourier transform to obtain the physical quantities in the time domain. Thereby, the expressions in Sec. \ref{secCoupling} are considered as discrete functions in relation to the wave frequencies.

The expressions are numerically calculated in the range of $100\,{\rm Hz}$ to $f_{GW_{\rm ISCO}}$ and in steps of $0.5\,{\rm Hz}$ in frequency. Numerical tests show that this frequency step is adequate to show the behaviors of the functions. The quantities of interest depend on the GW-EMW frequency as well as the resonant term given by $\omega_{MHD}-\omega_{GW}=\Delta \omega\rightarrow 0$. We use $\Delta \omega= 10^{-1}\,{\rm Hz}$ as a resonance condition between the waves. Numerical tests show that this value is adequate to represent the condition $\Delta \omega\rightarrow 0$.

\begin{figure}[ht]
\includegraphics[scale=0.42]{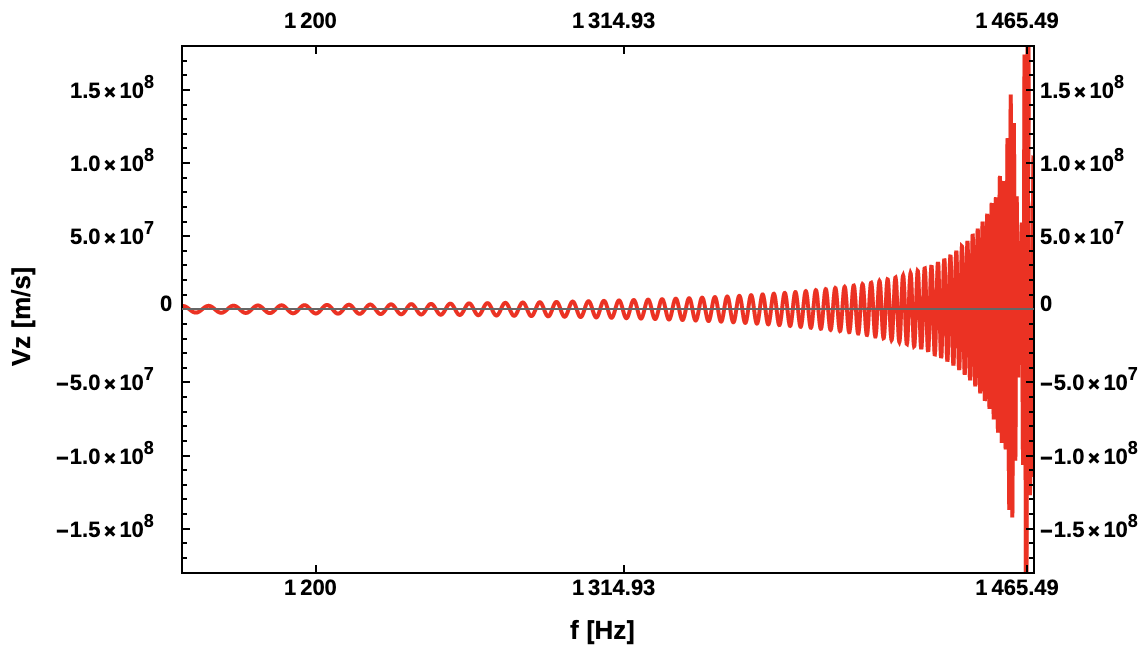}
\caption{\label{fig:VzxF} The evolution of the perturbed plasma velocity ($v_{z}^{(1)}$) as a function of $f_{GW}$. The plot is shown from the frequency of $1\,{\rm  kHz}$ as the interaction becomes more significant for $f_{GW} > 1.314\,{\rm Hz}$, that is, $2\,{\rm ms}$ before the merger.}
\end{figure}

The gravitational radiation propagates through the magnetized plasma and the charged particle trajectories are perturbed. According to equations in the Sec. \ref{secCoupling}, the plasma velocity increases with the GW amplitude (see Fig.~\ref{fig:VzxF}). Moreover the plasma is frozen in the magnetic field lines and it is in corotation with the stars.

\begin{figure}[ht]
\includegraphics[scale=0.42]{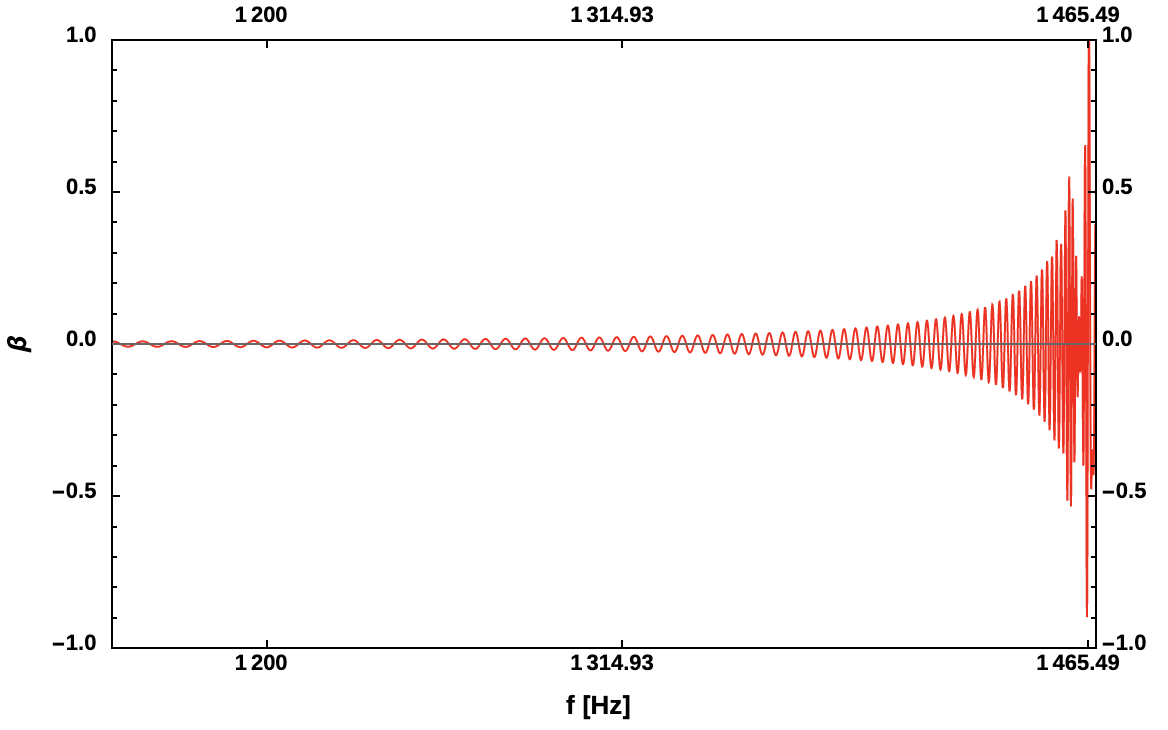}
\caption{\label{fig:betaxF} The evolution of $\beta$ in the rest frame as a function of $f_{GW}$. The star velocities and the perturbed plasma velocity due to the GW-EMW interaction are considered in the $\beta$ parameter.}
\end{figure}

The total plasma velocity in the laboratory frame is given by $v_{z}^{(1)}+v$, where $v$ is defined in Eq.~(\ref{eqnVelOrb}). In Fig.~\ref{fig:betaxF}, we can see the evolution of the $\beta$ parameter during the inspiral phase.

Instants before the merger, the plasma reaches velocity close to the speed of light and so the Lorentz factor goes to high values. We consider $\theta=\pi/4$ for the angle formed between the background magnetic field and the GW propagation. The phase velocity $u_f$ (fast MSW) does not depend on the $\theta$ angle. Additionally, the value $\pi / 4$ is the one that best matches the results obtained by \cite{64}.

From Figs.~\ref{fig:VzxF} and \ref{fig:betaxF} it is possible to verify that the GW-EMW interaction becomes more prevalent for frequencies $f_{GW} > 1.314\,{\rm kHz}$, which correspond to $\tau = 2\,{\rm ms}$ before the merger of the stars occurs. This result can be seen through the simulations developed by \cite{64}.

\begin{figure}[ht]
\includegraphics[scale=0.42]{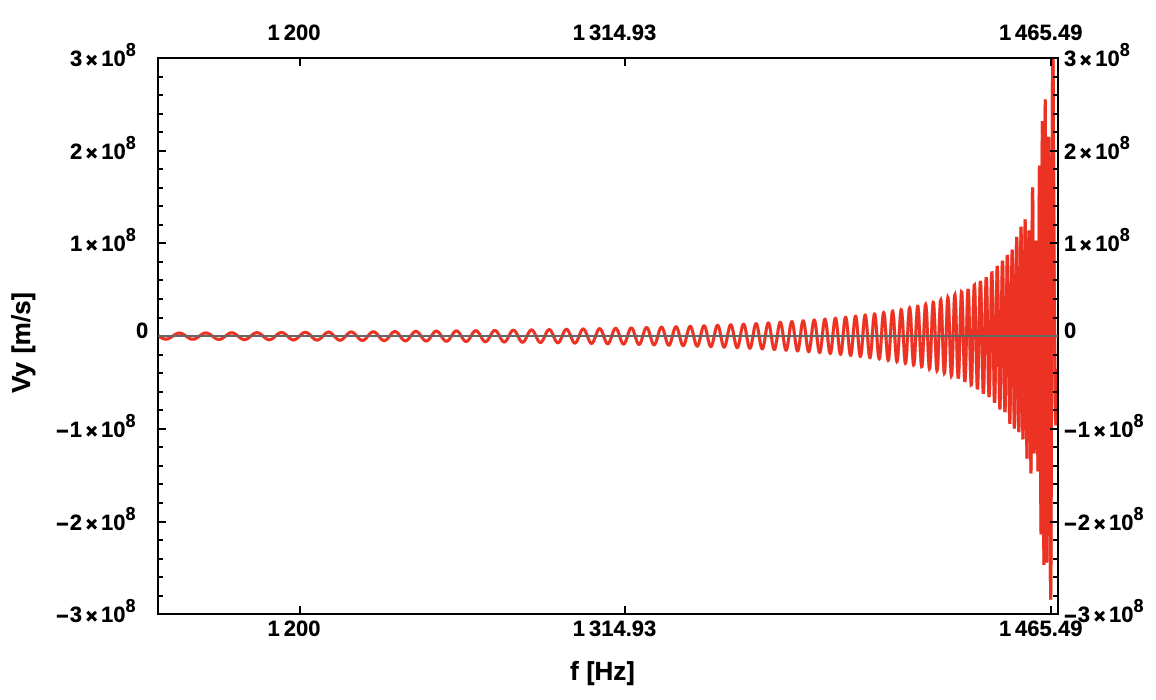}
\caption{\label{fig:VyxF} The evolution of the perturbed plasma velocity in the $y$-direction as a function of $f_{GW}$. The $v^{(1)}_y$ waveform is similar to the $v^{(1)}_z$.}
\end{figure}

For the Alfv\'en mode there is also the component $v^{(1)}_y(k,\omega)$ for the perturbed plasma velocity. Considering $\theta=\pi/4$, the value for $v^{(1)}_y(k,\omega)$ is slightly higher than $v^{(1)}_z(k,\omega)$ (see Fig. \ref{fig:VyxF}). This happens because the last one is proportional to $\sin^2(\theta)$ while the first one is proportional to $\sin(\theta)$.

The composition of the perturbed velocity components can not exceed the speed of light. In Fig.~\ref{fig:vstarsxF}, the maximum velocity due to the corotation of the stars is $1.18\times10^8\,{\rm m\,s^{-1}}$ ($\simeq0.395\,c$). Thus the maximum velocity due to GW-EMW interaction can not exceed $1.8\times10^8\rm{m\,s^{-1}}$ ($\simeq0.6\,c$). These limits are respected in our model.

The simulations developed by \cite{64} show that the maximum of the magnetic field in its poloidal and toroidal components increases significantly around $2\,{\rm ms}$ before the merger occurs. In particular, the maximum of the poloidal component increases by a factor of $\sim 3$ while the maximum of the toroidal component of the magnetic field, insignificant at the beginning of the inspiral phase, increases by a factor of $\sim 10$ for the merger. That is, the toroidal component increases from $10^{8}\,{\rm T}$ to $10^{9}\,{\rm T}$ during the last $2\,{\rm ms}$ for the merger. In the case of our model, this corresponds to sweep the band at a frequency of $1.314\,{\rm kHz}$ ($\tau = 2\,{\rm ms}$) to $1.465\,{\rm kHz}$ (ISCO).

In our model it is possible to see this effect in two different ways: either the magnetic field increases directly or the current densities are responsible for amplifying the magnetic field. In the first case, $B^{(1)}_x(k,\omega)$ and $B^{(1)}_y(k,\omega)$ are components of the toroidal magnetic field and they can reach high values at the coalescence. In the second case, $j^{(1)}_x(k,\omega)$ and $j^{(1)}_y(k,\omega)$ produce a magnetic field in the $z$-direction which is the the poloidal component while $j^{(1)}_z$, that lies in the $x-y$ plane, corresponds to the  toroidal component. 

See Eqs.~(\ref{eqndensitiesAW}) and (\ref{eqnjpboostmsw}). The current density $j^{(1)}_y\propto \, v^{(1)}_x/B^{(0)}_z$ where $B^{(0)}_z=B^{(0)}\cos(\theta)$ and $v^{(1)}_x(k,\omega)\propto 1/\tan(\theta)$. Thereby, $j^{(1)}_y \propto \, (B^{(0)}\sin(\theta))^{-1}$ and it increases when $\theta \rightarrow 0$. Additionally, see through Eqs.~(\ref{eqnVLaplaceBoostCoal_2}) that $v^{(1)}_x$ depends on the value of $v^{(1)}_z$. Close to the merger both $\gamma$ and the component $v^{(1)}_z$ reach high values (see also Fig. \ref{fig:VzxF}). All of these combinations can increase the value of $j^{(1)}_y$ close to the coalescence of the stars.

The current density $j^{(1)}_x(k,\omega)$ has two strong dependencies. The first relates to the term $(1-\beta u_{A_{\parallel}})^2/(u_{A_{\parallel}}-\beta)^2-1$ which decreases as $\beta$ increases. Note that in the ultrarelativistic regime, we have $\beta\rightarrow1$ and that term is approaching zero. The second is associated with $\gamma$ and the electromagnetic fields $E^{(1)}_{x}$ and $B^{(1)}_{y}$ which can reach high values during the merger and, thus, $j^{(1)}_x$ can increase in the final stages of evolution of the system.

In summary, the current densities acting on the $x-y$ plane can contribute to the increase in the value of the poloidal magnetic field. However, \cite{64} show that the growth of this component is not as significant until the merger of the stars occurs. After the merger, and until the collapse of the HMNS produces a black hole, the initial magnetic field grows as a result of the Kelvin-Helmholtz instability \cite{106,107}.

On the other hand, the toroidal magnetic field grows significantly during the coalescence. It is produced through the quantities $j^{(1)}_z(k,\omega)$, $B^{(1)}_x(k,\omega)$, and $B^{(1)}_y(k,\omega)$. Note that $j^{(1)}_z$ increases proportionally to the perturbed electric field $E^{(1)}_z$ which in turn is proportional to the perturbed velocity $v^{(1)}_x(k,\omega)$. Moreover, this current density gets very high close to the merger where $\beta\rightarrow1$.

\begin{figure}[ht]
\includegraphics[scale=0.42]{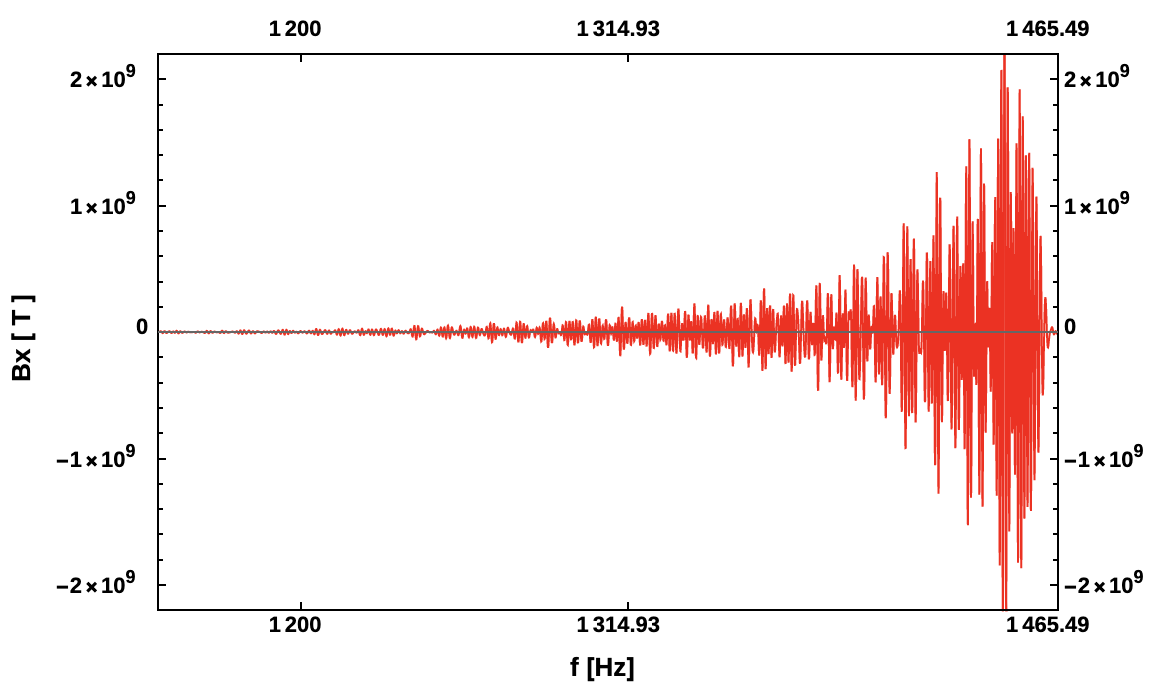}
\quad
\includegraphics[scale=0.42]{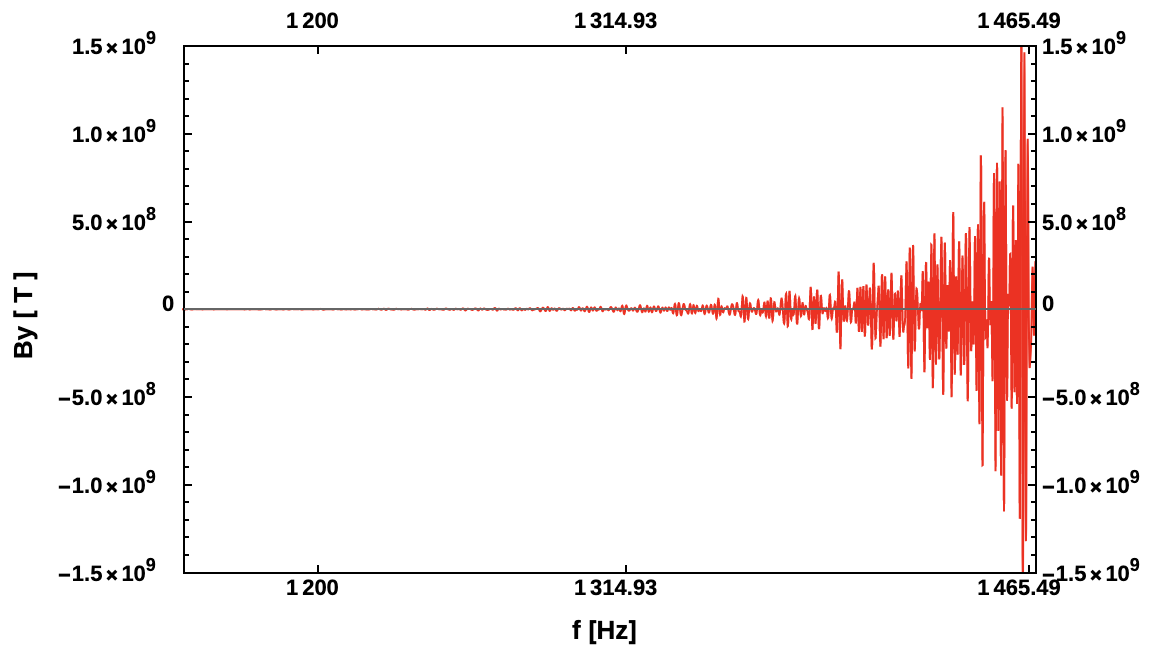}
\caption{\label{fig:BxF} The evolution of the maximum for the perturbed magnetic fields as function of $f_{GW}$. The upper panel shows the $B^{(1)}_x$ amplitude for the magnetosonic mode. The lower panel shows the $B^{(1)}_y$ amplitude for the Alfv\'en mode. We consider $\theta=\pi/4$ on both panels. The evolution of these magnetic field components, especially in the last $2\,{\rm ms}$ of the inspiral phase, are similar to those obtained from the simulations of \cite{64}.}
\end{figure}

The evolution of the maximum for the perturbed magnetic fields $B^{(1)}_x$ and $B^{(1)}_y$ is shown in Fig.~\ref{fig:BxF}. We can see that these magnetic components increase along the inspiral phase growing from $10^8\,{\rm T}$ up to $10^9\,{\rm T}$ in less than $2\,{\rm ms}$. This behavior is similar to that obtained by \cite{64}. Figure~\ref{fig:BxF} is shown for $\theta = \pi / 4$.

While the NSs orbit each other, gravitational radiation is emitted and it crosses  in the magnetized medium that surrounds the stars. The plasma  is frozen in the magnetic field lines and it extends up to the light cylinder radius ($\sim 10$ times the orbital radius).

The particles of the plasma are submitted for forces that change their trajectories. Thus, the GWs can excite the plasma parameters by transferring energy through that excitation. Equations~(\ref{eqnWBAW}) and (\ref{eqnWBMSW}), written in the frequency domain, determine the energy for the MHD modes, respectively, AW and MSW, excited by the GWs.

\begin{figure}[ht]
\includegraphics[scale=0.62]{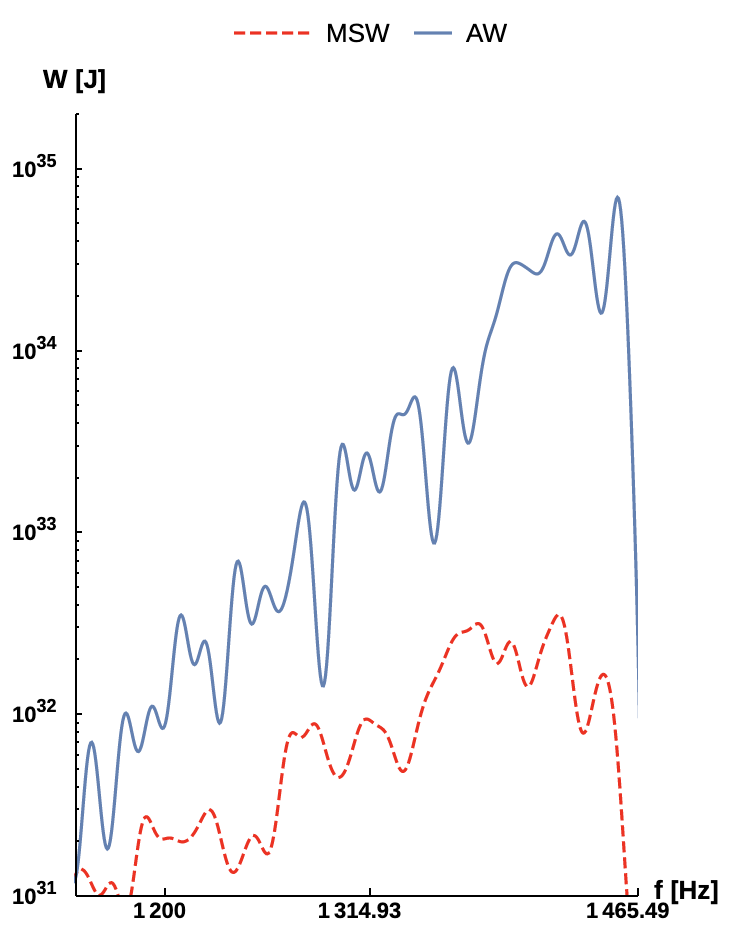}
\caption{\label{fig:WxF} The evolution of the electromagnetic energy excited by the GW-EMW interaction as function of $f_{GW}$. The red dashed curve represents the energy transferred to the magnetosonic mode through the GWs. The blue curve represents the energy transferred to the Alfv\'en mode through the GWs. The results presented here are similar to those found in the full GRMHD simulations developed by \cite{64}. We consider $\theta=\pi/4$ (see the text).}
\end{figure}

Considering that the GW propagation direction does an angle $\theta=\pi/4$  with the background magnetic field then, we can determine the energy transferred for the plasma during the inspiral phase (from $100\,{\rm Hz}$ up to $1465\,{\rm Hz}$). Figure~\ref{fig:WxF} presents the results for AW and MSW modes.

Note that the energy magnitudes obtained from Eqs.~(\ref{eqnWBAW}) and (\ref{eqnWBMSW}) are similar to those found in the simulations of \cite{64}. Furthermore, the electromagnetic energy increases between two and three orders of magnitude during the last $2$ ms ($\sim 1200 - 1465\,{\rm Hz}$) of the inspiral phase, a result that can be also seen in the full GRMHD simulation developed by \cite{64}. As we choose $\theta=\pi/4$, the electromagnetic energy is not the maximum, because of the term $\sin^2(\theta)$ in the energy equation.

However, for $\theta=\pi/4$, we can see the behavior of the Alfv\'en mode in Fig.~\ref{fig:WxF}. Although this mode is not fully in coherence with the GWs, we have for the Alfv\'en phase velocity $u_{A_{\parallel}}=u_{A}\cos(\theta) \approx 0.7\,c$ and this can be understood as ``an approximated condition of coherence for the GW-AW interaction''.

If we choose $\theta\rightarrow\pi/2$, then only the magnetosonic mode can be considered in the GW-EMW resonant interaction. The Alfv\'en phase velocity vanishes and this MHD mode is not more in coherence with the GWs. We expect that the maximum for the perturbed magnetic field and, therefore, the maximum transfer of energy to occur under resonant conditions. That is, with the waves in coherence.

\begin{figure}[ht]
\includegraphics[scale=0.56]{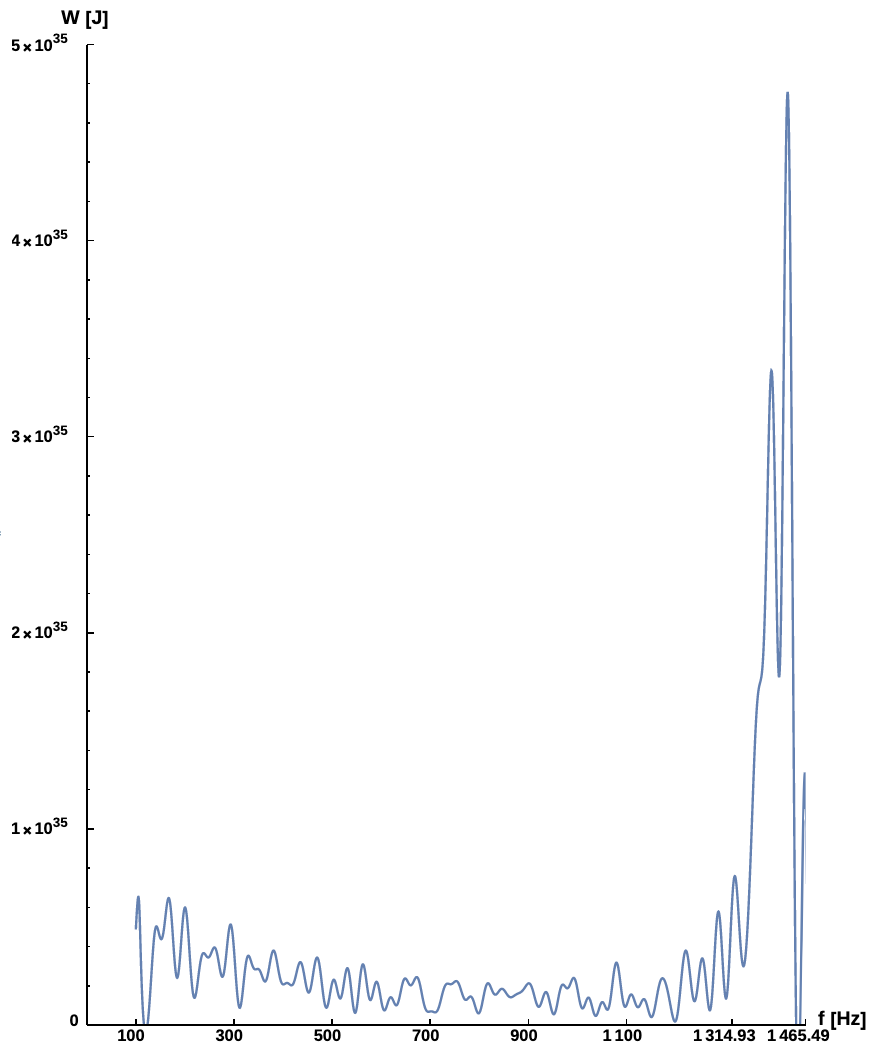}
\caption{\label{fig:WxFmax} The evolution of the perturbed electromagnetic energy for the MSW mode as a function of $f_{GW}$. The background magnetic field is perpendicular to the GW propagation direction, that is $\theta=\pi/2$. The energy is maximum very close to ISCO and can reach $\sim 4.5\times 10^{35}\,{\rm J}$ ($\sim 4.5\times 10^{42}\,{\rm erg}$). The MSW mode remains in coherence with the GWs for a wide range of values of the $\theta$ angle. In the particular case, $\theta =\pi / 2$, we have the maximum amplitude for the excitation of the electromagnetic energy by the MSW mode.}
\end{figure}

Applying $\theta=\pi/2$ in Eq. ~(\ref{eqnWBMSW}), we verify that the GW-EMW interaction, very close to ISCO, produces $\sim 4.5\times 10^{35}$ J ($\sim 4.5\times 10^{42}$ erg) for the MSW mode. This result is presented in Fig.~\ref{fig:WxFmax}.

Based on the previous discussions, we are in a position to take the properties of the source GW170817 \cite{104}, summarized in Table \ref{tableDatas}, in order to calculate, through our model, the energy transferred by the GW-EMW interaction.

\begin{figure}[ht]
\includegraphics[scale=0.6]{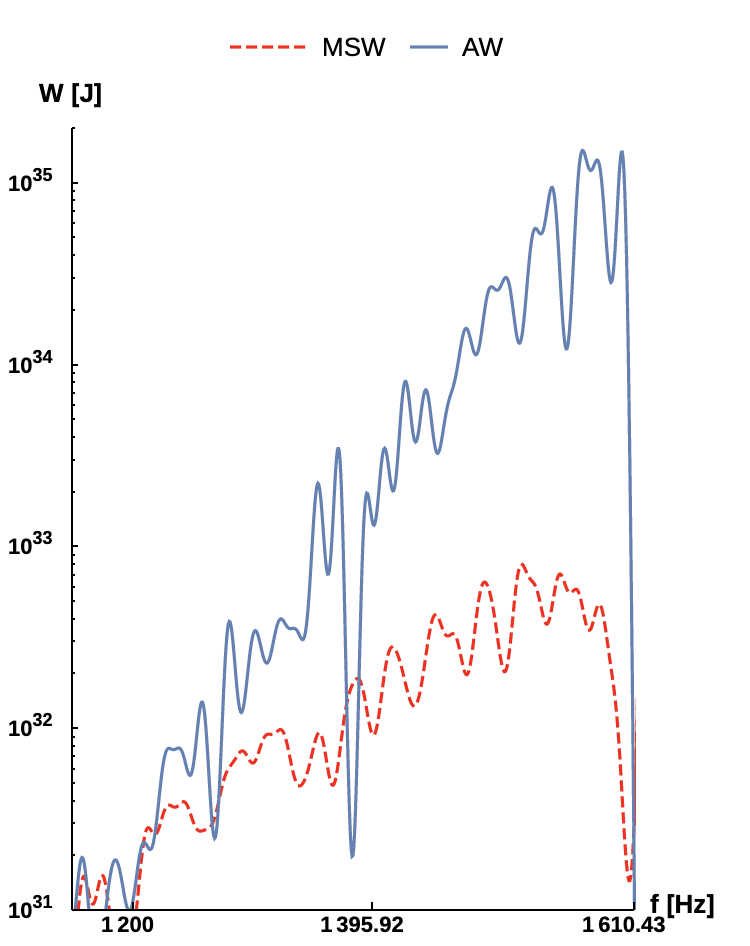}
\caption{\label{fig:WxFevent} The evolution of the electromagnetic energy excited by the GW-EMW interaction for the source GW170817 detected by aLIGO-VIRGO \cite{104}. The red dashed curve represents the energy transferred to the magnetosonic mode through the GWs. The blue curve represents the energy transferred to the Alfv\'en mode through the GWs. We consider $\theta=\pi/4$ (see the text).}
\end{figure}

Figure~\ref{fig:WxFevent} presents the evolution of the energy (AW and MSW modes) and considering $\theta=\pi/4$. The coalescence frequency for GW170817 is higher than the first case studied in this work. However, the evolutions of the two cases are very similar. The energy transferred to the plasma begins to significantly increase in the last $2\,{\rm ms}$ for the merger. In particular, close to ISCO, the transferred energy reaches its maximum value. 

\begin{figure}[ht]
\includegraphics[scale=0.5]{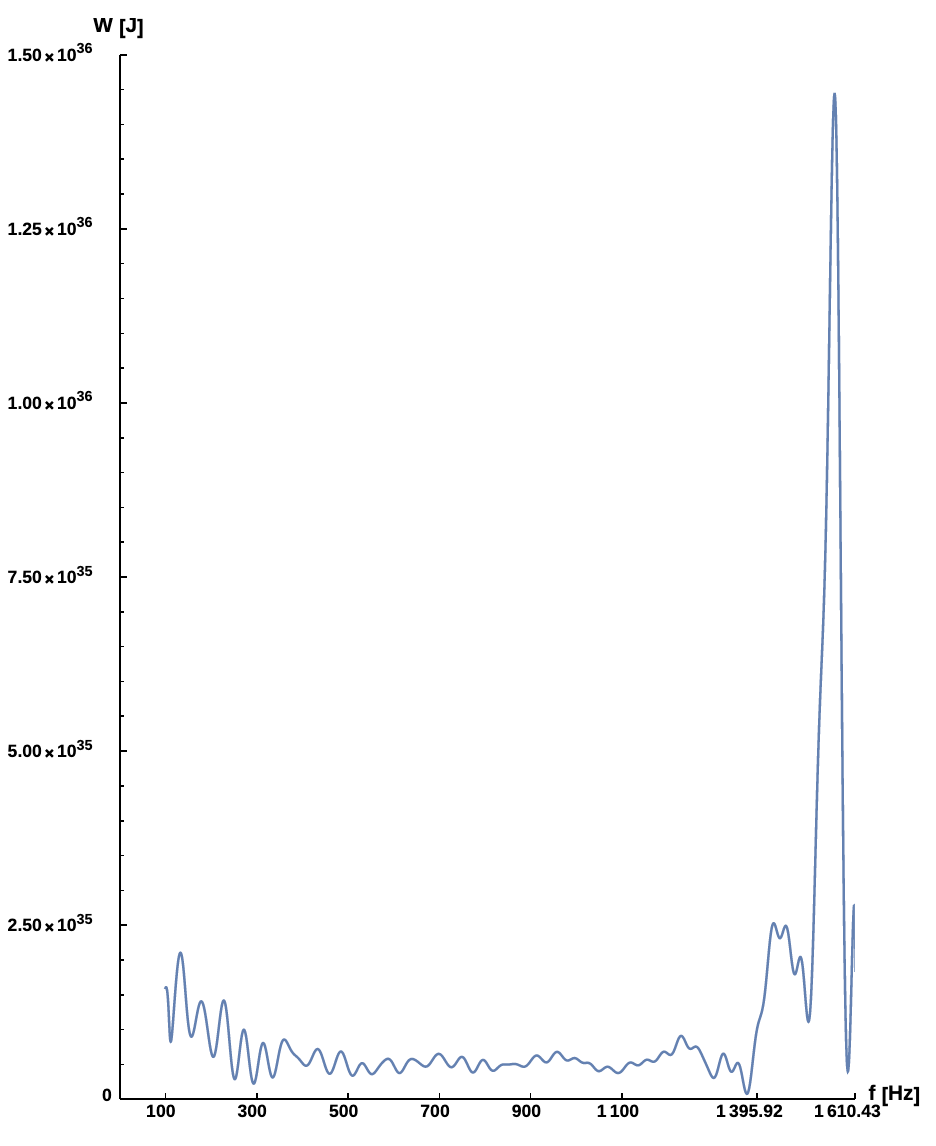}
\caption{\label{fig:WxFeventmax} The evolution of the perturbed electromagnetic energy for the MSW mode as function of $f_{GW}$ and associated with the source GW170817. The background magnetic field is perpendicular to the GW propagation direction ($\theta=\pi/2$). The energy is maximum close to ISCO and can reach $\sim 1.4 \times 10^{36}\,{\rm J}$ ($\sim 1.4\times 10^{43}\,{\rm erg}$).}
\end{figure}

Considering that the NSs of the event GW170817 had magnetic fields on their surfaces $\sim 10^{8}\, {\rm T}$ at the beginning of the inspiral phase, and, with $\theta = \pi / 2$, then the energy associated with the magnetosonic mode could reach values as high as $1.4 \times 10^{36}\,{\rm J}$ ($1.4 \times 10^{43}\,{\rm erg}$) before the merger. In Fig.~\ref{fig:WxFeventmax} we show the behavior of the energy as a function of the GW frequency. 

The resonant condition for the Alfv\'en mode occurs to $\theta=0$. According to the behavior of the MHD energies presented in this section, the energy of the Alfv\'en mode has amplitude higher than the magnetosonic mode for $\theta=\pi/4$.

As previously discussed, this value for the $\theta$ angle does not cause the GW-AW interaction to be in perfect coherence (resonance). Nevertheless, seeing that the Alfv\'en phase velocity is $u_{A_{\parallel}} \approx 0.7\,c$, then we consider $\theta=\pi/4$ as ``an approximated condition of coherence'' for the representation of the AW mode energy. 

Through the behavior of the electromagnetic energy excited by the MHD modes, as shown in Fig.~\ref{fig:WxFevent}, we can obtain the representation of the $W_{MSW}^{(B)}$ and $W_{AW}^{(B)}$ functions through a power series of the GW frequency (or in terms of the system's orbital frequency, since $f_{GW}= 2f_{orbital}$). That is, we can represent $W\propto f_{GW}^{p}$ and the result of that analysis produces

\begin{align}
\label{wmswn}
 W_{MSW}^{(B)}=&-0.338085 f^{11}+5315.82 f^{10} \nonumber \\
 &-3.6113\times 10^7 f^9+1.38831\times 10^{11}f^8 \nonumber\\
 &-3.32265\times 10^{14}f^7+5.13136\times 10^{17}f^6 \nonumber\\
 &-5.12839\times 10^{20}f^5+3.23774\times 10^{23} f^4 \nonumber\\
 &-1.22143\times 10^{26} f^3+2.48183\times 10^{28}f^2 \nonumber\\
 &-2.23765\times 10^{30} f+6.56468\times 10^{31} \ ,
 \end{align}
and
 
 \begin{align}
 \label{wawb}
 W_{AW}^{(B)}=&-0.0579517 f^{12}+1006.29 f^{11} \nonumber\\
 &-7.6465\times 10^6 f^{10}+3.34092\times 10^{10}f^9 \nonumber\\
 &-9.27454\times 10^{13} f^8+1.70623\times 10^{17} f^7 \nonumber\\ 
 &-2.10551\times 10^{20}f^6 +1.72578\times 10^{23} f^5 \nonumber\\
 &-9.10102\times 10^{25} f^4+2.90493\times 10^{28}f^3\nonumber \\
 &-5.02533\times 10^{30} f^2+3.79387\times 10^{32}f \nonumber\\
 &-7.20951\times 10^{33} \ .
\end{align}

In \cite{59} full GRMHD simulations of NS binaries are presented. One of their results shows the behavior of the luminosities as functions of time. For the simulated systems, the luminosity can be represented in terms of powers of the orbital frequency of the binary system.

In particular, considering that the magnetic field on the surface of stars remains constant during evolution, \cite{59} shows that $L\propto \Omega^p$. Looking at the simulations of those authors that present a magnetic configuration similar to that of our work, we can see that they present $p\approx 1-2$ for the beginning of the inspiral phase.

In the final moments for the merger the growth of luminosity is of the form $p\sim~ 12$. As luminosity and energy are directly proportional quantities, the expansion in power series expressed in our work through Eqs. (\ref{wmswn}) and (\ref{wawb}) show consistency with the simulations of   \cite{59}.

\section{Refractive Index}
\label{secRefIndex}

Uniform plane waves, $\omega$, can only propagate in plasmas with frequencies, $\omega_p$, such as $\omega>\omega_p=\sqrt{N_e\,e^2/(\epsilon_0 m_e)}$. Otherwise, it occurs in the total reflection of the wave \cite{108} within the domain of the binary system. A natural example is the interplay region formed between the interplanetary medium and Earth's ionosphere, although the electron density in the ionosphere increases gradually with the height. Reflections occur when $\omega<\omega_p$. 

The coefficients of reflection ($R$) and transmission ($T$) are given, respectively, by

\begin{subequations}
\label{eqnReflexaoPlasma}
\begin{align}
    R &=\frac{\sqrt{\omega^2-\omega_p^2}-\omega}{\sqrt{\omega^2-\omega_p^2}+\omega} \ , \\
    T &= \frac{2\sqrt{\omega^2-\omega_p^2}}{\sqrt{\omega^2-\omega_p^2}+\omega} \ .
\end{align}
\end{subequations}

For $\omega>\omega_p$, the values for $R$ and $T$ are real numbers and the plane wave $e^{i(\vec{k}\cdot \vec{r}-\omega t)}$ propagates through the conductive layer. Otherwise, for $\omega<\omega_p$, $R$ is an imaginary number and the wave does not propagate. It reflects in the conducting layer.

Another way to evaluate the coefficients of reflection and transmission is through the boundary condition of the electrodynamics in the linear medium. Consider a plane wave $\vec{E}_I (z,t) = \vec{E}_{0_{I}} e^{i(kz-\omega t)}$ propagating in the $\hat{z}$-direction and it strikes at the interface separating two mediums with different refractive indexes ($n_1, n_2$). In terms of the amplitude of the incident wave, the reflected and transmitted wave amplitudes are

\begin{subequations}
\begin{align}
    E_{0_{R}} &= \bigg(\frac{1-\zeta}{1+\zeta}\bigg)E_{0_{I}} \ , \\
    E_{0_{T}} &= \bigg(\frac{2}{1+\zeta}\bigg) E_{0_{I}} \ ,
\end{align}
\end{subequations}
where $\zeta \equiv \mu_1 n_2/(\mu_2 n_1)$, $\mu_1$ and $\mu_2$ are the permeability of the medium 1 and 2, respectively.

Thus, the coefficients of reflection and transmission are given, respectively, by

\begin{subequations}
\label{eqnReflexao}
\begin{align}
    R &= \bigg(\frac{n_1-n_2}{n_1+n_2}\bigg)^2 \ , \\
    T &= \frac{4n_1n_2}{(n_1+n_2)^2} \ .
\end{align}
\end{subequations}

$ R $ and $ T $ measure the fraction of the incident energy that is reflected and transmitted, respectively. Note that $ R + T = 1 $ \cite{109}. Therefore, the goal is to find the refractive index of the plasma medium through the MHD waves that propagate in it.

First, we apply $\times\vec{B}^{(0)}$ in the momentum equation, Eq. ~(\ref{eqnGRM:8}), substituting $\vec{E}^{(1)}=-\vec{v}^{(1)}\times\vec{B}^{(0)}$ and using the vector property $\vec{A}\times(\vec{B}\times\vec{C})=\vec{B}(\vec{A}\cdot\vec{C})-\vec{C}(\vec{A}\cdot\vec{B})$. Thus,

\begin{equation}
-\frac{w^{(0)}}{c^2}\frac{\partial^2}{\partial t^2}\vec{E}^{(1)} + \frac{\partial}{\partial t} \vec{\nabla}p^{(1)}\times \vec{B}^{(0)} =- \frac{\partial}{\partial t}\vec{j}^{(1)}B^{(0)^{2}} \ ,
\end{equation}
with 

\begin{equation}
\frac{\partial}{\partial t} \vec{\nabla}p^{(1)}\times \vec{B}^{(0)} - \gamma p^{(0)}\vec{B}^{(0)}\times[\vec{\nabla}(\vec{\nabla}\cdot\vec{v}^{(1)})]=0 \ ,
\end{equation}
that is calculated from the pressure equation [Eq. ~(\ref{eqnGRM:5})]. Then,

\begin{align}
-\frac{w^{(0)}}{c^2}&\frac{\partial^2}{\partial t^2} \vec{E}^{(1)} -\gamma p^{(0)}\vec{\nabla}\times\vec{\nabla}\times\vec{E}^{(1)} \\ &+ \frac{\gamma p^{(0)}}{B^{(0)^2}}(\vec{B}^{(0)}\cdot\vec{\nabla})(\vec{\nabla}\cdot\vec{E}^{(1)})\vec{B}^{(0)} =- \frac{\partial}{\partial t}\vec{j}^{(1)}B^{(0)^2} \nonumber \ ,
\end{align}
with the analogue equations to the Amp\`ere and Faraday equations in Eq. ~(\ref{eqnGRM:2}). The term $\partial \vec{j}/\partial t$ can be isolated:

\begin{align}
    \label{eqnOEM_E}
    \Big(u_m^2-\frac{\partial^2}{\partial t^2}\Big)\vec{E}^{(1)} + \Bigg[\frac{c^2_s}{u_A^2}&(\vec{u}_A\cdot\vec{\nabla})\vec{u}_A^{\phantom{A}2}-u_m^2\vec{\nabla}\Bigg](\vec{\nabla}\cdot\vec{E}^{(1)}) \\ \nonumber &=u^2_A\bigg(\frac{\partial} {\partial t}\vec{j}_{E}+\vec{\nabla}\times\vec{j}_B\bigg) \ ,
\end{align}
where $u_A = cB^{(0)}/(\sqrt{\mu_0 w_{tot}})$ and $u_m^2 = c_s^2 + u_A^2$ are the Alfv\'en and compressional velocities, respectively.

Applying the Fourier transform and remembering of the expression for $\vec{j}_E$, then Eq. ~(\ref{eqnOEM_E}) can be algebraically calculated for $\omega/k$, which is the velocity of the electromagnetic wave produced in the plasma through the GW. That is,

\begin{align}
\label{eqnFouierOEM_E}
\begin{pmatrix}
\omega^2-k^2 u_m^2  &0  &-\frac{c_s^2 u_{A_{\parallel}}u_{A_{\perp}}k^2}{u_A^2} \\ 
\omega^2-k^2 u_m^2  &0  &0 \\ 
\omega^2-k^2 u_m^2  &0  &k^2\bigg(u_m^2-\frac{c_s^2u_{A_{\parallel}}^2}{u_A^2}\bigg)  \\ 
\end{pmatrix}&\begin{pmatrix}
E_x^{(1)}\\
E_y^{(1)}\\
E_z^{(1)}\\
\end{pmatrix} \\ \nonumber &=u_A^2B_x^{(0)}k\omega\begin{pmatrix}
 h_{\times}\\
 -h_{+}\\
 0\\
\end{pmatrix} \ .
\end{align}

Therefore, we note that the $\hat{x}$-component of the electric field depends on the $h_{\times}$ GW polarization. On the other hand, the $\hat{y}$-component depends on the $h_{+}$ GW polarization. From the equation of the $\hat{z}$-direction a relation for $\omega$ and $k$ can be obtained. This results in

\begin{equation}
    \frac{\omega}{k} = \pm \frac{c_s u_{A_\parallel}}{u_A} \ .
\end{equation}

The velocities of the electromagnetic waves depends on the plasma sound velocity, which normally does not present high values when compared to the speed of light. The refractive index is calculated as $n=ck/\omega$. Using the previous results and taking $c_s = \sqrt{\gamma p^{(0)}/w_{tot}} c$, $u_A = B^{(0)}c/\sqrt{\mu w_{tot}}$\,, and $u_{A_{\parallel}} = u_A \cos{\theta}$, where $\theta$ is the angle that the gravitational wave vector does with the background magnetic field, we have

\begin{align}
\label{eqnRefractiveIndexPlasma}
    n &= \frac{1}{c_s}\frac{1}{\cos{\theta}} \\ \nonumber &=\frac{1}{\cos{\theta}}\sqrt{\frac{w_{tot}}{\gamma p^{(0)}}} = \frac{1}{\cos{\theta}}\sqrt{\bigg(\mu^{(0)}+p^{(0)} +\frac{B^{(0)^2}}{\mu}\bigg)\frac{1}{\gamma p^{(0)}}} \ .
\end{align}

Thus, the refractive index depends on the plasma parameters: mass density, pressure, and magnetic field. This makes sense, since in a magnetized plasma system, the density, the pressure, and the magnetic field are the parameters responsible for its evolution. For the BNSs, the magnetic field is high, so the refractive index has also to be high.

In relation to the surrounding medium, the difference from a more refringent medium to another less refringent is that the former does not allow the perturbed electromagnetic radiation to escape from the system.

To understand this result, we can take Eq. ~(\ref{eqnReflexao}) with $n_1$ representing the plasma and $n_2$ the surrounding medium. As $n_1\gg n_2$, then we have $R\simeq 1$ and $T\rightarrow 0$.

Through the equation $\omega_p=\sqrt{N_e\,e^2/(\epsilon_0 m_e)}$, with the Goldreich-Julian density $\approx 10^{12} \rm{cm}^{-3}$ to the plasma surrounding the binaries \cite{98}, the MHD wave would need to have a higher frequency than $1$ GHz to cross the conducting layer. As the perturbed MHD wave is in coherence with the GW frequency, so reaching values $\sim \,{\rm kHz}$, then it collides with the conducting layer being totally reflected.

\section{The Poynting Vector}
\label{secPoyntingVector}

The energy flux density (energy per unit area and per unit time) carried by the electromagnetic fields can be determined by the Poynting vector. It is given by

\begin{equation}
    \vec{S}^{(1)} = \frac{1}{\mu}\vec{E}^{(1)}\times\vec{B}^{(1)} \ .
\end{equation}

It can be calculated for the two MHD modes: Alfv\'en and magnetosonic modes. The AW has an electric field in the $E^{(1)}_x\hat{x}$ and $E^{(1)}_z\hat{z}$ directions. The magnetic field is in the $B^{(1)}_y\hat{y}$ direction. Thus, $\vec{S}$ is in the $\hat{z}$ and $\hat{x}$ directions. In particular,

\begin{equation}
\vec{S}_{AW} = -\frac{1}{\mu}v_y^{(1)}B^{(1)}_y\vec{B}^{(0)} \ .
\end{equation}

Substituting the expressions from Sec. \ref{secCoupling}, Eqs. ~(\ref{eqnVLaplaceBoostCoal}) and ~(\ref{eqnBLaplaceBoostCoal}), for the rest frame, we obtain

\begin{equation}
    \label{eqnPoyntingAlfven}
    \vec{S}_{AW} \propto h_{\times}^{2}\vec{B}^{(0)}\frac{\omega^{2}}{\omega^2-k^2} \ ,
\end{equation}
since $\omega/k=c_s u_{A_{\parallel}}/u_A$. Based on the above equation, the transported energy depends strongly on the frequency and amplitude of the GWs.

For the MSW mode, the electric field is in the $E_y^{(1)}\hat{y}$ direction while the magnetic field is in the $B^{(1)}_x\hat{x}$ direction. In this case, the Poynting vector will be in the $\hat{z}$ direction. That is,

\begin{equation}
    \label{eqnPoyntingMSW}
    \vec{S}_{MSW}=-\frac{1}{\mu}E_y^{(1)}B_x^{(1)}\hat{z} \ .
\end{equation}

Using Eqs.~(\ref{eqnVLaplaceBoostCoal_2}) and~(\ref{eqnBLaplaceBoostCoal_2}), we find the strong dependence of the Poynting vector with the frequency and amplitude of the GWs that interact with the plasma.

Electromagnetic fields transport energy and momentum. When the light strikes a perfect absorber, it transmits its momentum to the surface. Thus, the radiation pressure can be obtained from the Poynting vector as

\begin{equation}
    P = \frac{S}{c} \propto (h_{+}^2 + h_{\times}^2)\, \omega^2 \ .
\end{equation}

Hence, for the model discussed here with the significant Poynting vector, especially at the instants close to the merger, the radiation pressure can reach high values.

An interesting analysis of the Poynting vector consists to find the plasma parameters that amplify or attenuate the energy carried by the MHD modes. Thus, taking the first derivative of the Poynting vector with respect to time, we have

\begin{equation}
\frac{\partial}{\partial t} \vec{S} = \frac{1}{\mu}\bigg[\frac{\partial}{\partial t}\vec{E}^{(1)}\times\vec{B}^{(1)}+\vec{E}^{(1)}\times\frac{\partial}{\partial t}\vec{B}^{(1)}\bigg]  \ .
\end{equation}
 
Substituting the expressions of $\vec{B}^{(1)}$ and $\vec{E}^{(1)}$, we obtain

\begin{widetext}
\begin{equation}
\label{eqnDerivativePoynting}
    \mu\varepsilon\frac{\partial}{\partial t}\vec{S} = \frac{1}{\mu}\bigg[-\frac{1}{2}\vec{\nabla}B^{(1)^2} + (\vec{B}^{(1)}\cdot\vec{\nabla})\vec{B}^{(1)}\bigg]+\varepsilon\bigg[-\frac{1}{2}\vec{\nabla}E^{(1)^2}+(E^{(1)}\cdot\vec{\nabla})E^{(1)}\bigg]
    -\bigg(\frac{1}{\mu}\vec{j}_E\times\vec{B^{(1)}} +\varepsilon\vec{E}^{(1)}\times\vec{j}_B \bigg) \ .
\end{equation}
\end{widetext}

The first two terms on the right-hand side correspond to the Maxwell stress tensor \cite{109}. The derivatives of the Poynting vector and the Maxwell stress tensor are equal to the force per unit of volume on the electrical charges in the plasma. Therefore, the interaction of the GWs with the electric and magnetic fields exerts the role of force acting on the charges contained in a certain volume. This result corroborates the assertion that the coupling of GWs with magnetized plasma provides energy and momentum to the charges of the plasma. Finally, if the derivative of $\vec{S}$ is zero, which corresponds to the condition of maximum energy transport, then the force applied on the charges should be equal to the Maxwell stress tensor.

\section{Final Remarks}\label{remarks}

The main objective of this work is to present a semianalytical formalism, which allows to study the coupling between GWs with strongly magnetized plasma presents in BNS systems. With the detection of the first BNS merger observed in GWs (GW170817) followed by electromagnetic counterpart (GRB 170817A), the era of the multimessenger astronomy began.

Thus, it becomes important to develop not only numerical simulations in full GRMHD but also to develop semianalytical tools that allow exploring the space of parameters, contributing to our understanding on the physics of compact objects and the physical mechanisms associated with the generation of sGRBs.

Our work has deepened previous studies developed mainly in \cite{65,69,70,71,72,75,76,77,78}. However, none of these works presented the explicit mathematical calculations as we developed in this article. Additionally, in these previous works, the authors assume a delta function to calculate the interaction and the transference of energy between the GWs and the magnetized plasma of the binary system. We present the expressions for the full coalescence phase and discuss the consequences of this interaction.

We consider that the plasma is in the ideal MHD approximation and can interact with the GW released by the coalescence of the stars. While the NSs orbit each other, the system releases gravitational radiation that crosses the magnetized medium surrounding the stars. The plasma is frozen in the magnetic field lines extending up to the light cylinder radius ($\sim 10$ times the orbital radius).

Plasma particles are forced by the $h_{+}$ and $h_{\times}$ amplitudes to change their trajectories and as a result the plasma is excited by the propagation of the GWs. As a direct consequence of the interaction with the GWs, two wave modes are excited in the plasma. Initially, the set of equations is  written in the comoving frame but we need to study the consequences of the coupling in the laboratory frame. 

Once all of our formalism was developed and presented, we apply it to two different cases: (a) for the computational simulations in full GRMHD of \cite{64} and (b) for the detected event GW170817 \cite{104}. For the first case, we calculated the energies associated with the AW and MSW modes and considering $\theta=\pi / 4$. It is important to emphasize that the MSW mode remains in coherence with the GWs over a wide range of $\theta$ values while the AW mode remains in coherence only for $\theta = 0$. 

If $\theta = \pi / 4$, the Alfv\'en phase velocity is $\sim 0.7 \, c$ and there is no perfect GW-AW resonance. We denominate this as ``approximated condition of coherence'' and it is permitted to obtain an estimate of how the energy associated to the AW mode behaves. For the MSW mode we also use $\theta = \pi / 2$ which provides the maximum amplitude for the excitation of the electromagnetic energy.

Concerning the case (a), we calculate the perturbed quantities for the BNS and compare the results with the simulations of \cite{64}. We follow the system evolution with our modeling  from $100\,{\rm Hz}$ until the ISCO frequency of the system. Figures~(\ref{fig:VzxF}), (\ref{fig:VyxF}), and (\ref{fig:BxF}) show the evolution of the main physical quantities.

We calculate the electromagnetic energy transferred by the MHD modes for two angles ($\theta=\pi/4$ and $\theta = \pi/2$). As discussed above, the magnetosonic mode remains in coherence for both angles. However, the Alfv\'en mode is only in coherence with the GWs when $\theta\rightarrow0$.

Figure~(\ref{fig:WxF}) shows the electromagnetic energy excited when $\theta=\pi/4$. The Alfv\'en mode reachs  $\sim 10^{35}\,{\rm J}$ before merger, while the magnetosonic mode reaches $\sim 10^{32}\,{\rm J}$ at the ISCO frequency. However, for $\theta=\pi/2$, we consider only the MSW mode and the electromagnetic energy is $\sim 4.5\times 10^{35}\,{\rm J}$ for frequencies very close to ISCO (see Fig.~\ref{fig:WxFmax}).

We compare these results with those obtained by \cite{64} and, in general, our results (e.g, electromagnetic energy modes and magnetic fields) have similar evolution, especially in the last $2\,{\rm ms}$ for the merger. 

The second case study is related to the source GW170817 detected by the LIGO-VIRGO collaboration \cite{104}. The main parameters are: $2.73\, M_{\odot}$ for the total mass and $M_{c}=1.186\,M_{\odot}$ for the chirp mass. On the other hand, we consider for each star $B^{(0)}=10^{8}\,{\rm T}$ and $R_{\star}=1.36\times10^4\,{\rm m}$ as typical parameters for, respectively, the magnetic field on the star surfaces and for their radii \cite{59,64}. Using our formalism it is then possible to obtain $f_{GW_{\rm ISCO}}\simeq1610\,{\rm Hz}$ which produces $\beta=0.035 f^{1/3}_{GW}$. 

We find that the total stored energy in the plasma resulting from coupling with the GWs can reach maximum value $\sim 10^{33}\,{\rm J}$ for the MSW and $\sim 10^{35}\,{\rm J}$ for the AW when $\theta=\pi/4$. We show in Fig. \ref{fig:WxFevent} the behaviors of the energies of the MHD modes with the frequency during the inspiral phase until the ISCO frequency. For $\theta=\pi/2$ the MSW mode can store $\sim 10^{36}\,{\rm J}$ instants before the merger as can be seen in Fig.~\ref{fig:WxFeventmax}.

The magnetosonic mode is more efficient, since its perturbation is perpendicular to the background magnetic field as a compressional wave. This is the same condition to an efficient coupling between the GWs and the plasma waves. The shear wave excites parallel oscillations to $\vec{B^{(0)}}$, reducing the stored energy.

The stored magnetic energy evolves with the GW frequency. Thus, lower frequencies also contribute to store the energy although the most important contribution occurs for instants immediately before the merger, as can be seen in Figs.~\ref{fig:WxF} and \ref{fig:WxFevent}.

The MHD waves can propagate through in the magnetized plasma. Nevertheless, we need to assess whether these waves can flow out of the region in which they are generated. The important parameter in this case is the refractive index. We show that this parameter depends on the mass density, pressure, and magnetic field, Eq.~(\ref{eqnRefractiveIndexPlasma}). For the BNSs, the magnetic field is strong, so the refractive index can reach huge values.

On the other hand, the medium surrounding the plasma has refractive index  $\sim 1$. The difference from a more refringent medium, that means the plasma around the BNS, to another less refringent (the interstellar medium) is that the perturbed electromagnetic radiation can not escape the system.

Uniform plane waves, $\omega$, can only propagate in plasmas with frequencies such as $\omega>\omega_p=\sqrt{N_e\, e^2/(\epsilon_0 m_e)}$. Considering the Goldreich-Julian density $\approx 10^{12} \rm{cm}^{-3}$ for the plasma surrounding the binaries \cite{98}, the MHD waves would need to reach frequencies higher than $1\,{\rm GHz}$ to be able to cross the conducting layer of the plasma. Since the perturbed MHD wave be in coherence with the GW frequency, values $\lesssim 1\,{\rm kHz}$, it collides with the conducting layer being totally reflected and remaining confined in the system.

We show in Eqs.~(\ref{eqnPoyntingAlfven}) and ~(\ref{eqnPoyntingMSW}) that the energy depends strongly on the frequency and amplitude of the gravitational perturbation. When a wave strikes a perfect absorber, it transmits its momentum to the surface. Our calculations show that the radiation pressure is $P \propto (h_{+}^2 + h_{\times}^2)\, \omega^2$.

During the coalescence, the frequency and amplitude increase until they reach the highest values in the merger of the system. Therefore, the radiation pressure also reaches huge values. Theses characteristics are important to understand the phenomenons that arise with BNSs, primarily, the bursts and the high Lorentz Factor in the sGRBs.

Regarding the GRB 170817A, it was considered as possibly being an sGRB due to its duration of $(2\pm 0.5)\,{\rm s}$. However, the equivalent isotropic energy is $(5.35\pm 1.26)\times 10^{39}\,{\rm J}$, which results in a value lower by three orders of magnitude than the weakest sGRB known \cite{105}. It is not clear what the origin of this ``weakness'' is. However, some authors have argued that the weak emission is consistent with an off-axis viewing effect \cite{110,111}.

Our work shows that for $\theta = \pi / 2$ the MSW mode could reach energies $\sim 10^{36}\,{\rm  J}$. As this MHD mode remains in coherence with the GWs for a wide range of values ​​of $\theta$, the resonance condition between the GW-MHD is preserved. The energies inferred for the event GW170817 could be reached in our model if $B_{\star}\sim  2\times 10^{9}\,{\rm T}$. Even greater energies could be obtained by increasing the initial magnetic field on the surface of the stars.

Another point to consider is that our formulation, in its current state, allows us to consistently follow the evolution of the system during the inspiral phase until very close to the ISCO. The subsequent phase in which the formation of HMNS occurs and the final collapse to form a black hole, as shown in \cite{64}, are not followed in our model.

In particular, \cite{64} argue that the magnetorotational instability could be generated about $5 \, {\rm ms}$ after the formation of the black hole. Thus, a significant amplification of the poloidal and toroidal magnetic fields can occur. This can contribute to increasing the energy of the MHD modes leading to values ​​much higher than $10^{39}\,{\rm J}$.

We show that GWs can coherently excite MHD waves that, in turn, carry energy and momentum. Therefore, the GW-MHD coupling mechanism could become an important player for studying the engine associated to the generation of sGRBs.

Our results have consistency with the evolution of the physical parameters shown in the full GRMHD simulations of \cite{64} and with the polynomial expressions to relate energy with frequency as discussed in \cite{59}. This can be seen through our Eqs.~(\ref{wmswn}) and (\ref{wawb}).

At last, the gravity participates as the fundamental force to lead the coalescence of the stars in a binary system. Furthermore, the GWs can be part of a more fundamental mechanism to help produce the gamma-ray bursts and so to accelerate the baryonic matter for high Lorentz factor.

\begin{acknowledgments}
A.S.G. was financed in part by the Coordena\c c\~ao de Aperfei\c coamento de Pessoal de N\'ivel Superior (CAPES) - Finance code 001. O.D.M. would like to thank the Brazilian Agency CNPq for partial financial support (Grant No. 303350/2015-6). We  would also like  to  thank  the anonymous  referee for his/her critical comments and suggestions, which have helped to improve the manuscript.
\end{acknowledgments}

\appendix

\section{\label{apen} Terms $\Xi$ and $\Pi$ of the perturbed magnetic field}

Equation~(\ref{eqnBLaplaceBoostCoal_2}) presents the perturbed magnetic field by MSWs. The terms $\Xi$ and $\Pi$ are, respectively,

\begin{widetext}
\begin{align}
\Xi(k,\omega) \equiv \frac{B_{x}^{(0)}}{1-\beta} \Bigg\{&\frac{u_{s}}{\omega^2-k^2u_{s}^2}\bigg[\omega\frac{u_{s}-\beta}{1-\beta u_{s}}+\frac{k u_{s}}{(1-\beta u_{A_{\parallel}})^2} \Big[(u_{A_{\parallel}}-\beta)^2+\frac{u_{A_{\perp}}^2}{\gamma^2}\Big]\bigg]
\\
-&\frac{u_{f}}{\omega^2-k^2 u_{f}^2} \bigg[\omega\frac{u_{f}-\beta}{1-\beta u_{f}}+\frac{k u_{f}}{(1-\beta u_{A_{\parallel}})^2}\Big[(u_{A_{\parallel}}-\beta)^2+\frac{u_{A_{\perp}}^2}{\gamma^2}\Big]\bigg] 
\nonumber \\
&+2\frac{(u_{s}-\beta)^2(u_{f}-\beta)^2}{(1-\beta u_{f})^2 (1-\beta u_{s})^2} \cdot
\nonumber \\ 
\Bigg[&\frac{u_{s}}{\omega^2-k^2u_{s}^2}\frac{1-\beta u_{s}}{(u_{s}-\beta)^2}\bigg[\omega\frac{u_{s}-\beta}{(u_{A_{\parallel}}-\beta)^2}\Big[1-\frac{u_{A_{\perp}}^2}{\gamma^2(1-\beta u_{A_{\parallel}})^2}\Big]+k u_{s}\frac{1-\beta u_{s}}{(1-\beta u_{A_{\parallel}})^2}\bigg]
\nonumber \\
-&\frac{u_{f}}{\omega^2-k^2u_{f}^2}\frac{1-\beta u_{f}}{(u_{f}-\beta)^2}\bigg[\omega\frac{u_{f}-\beta}{(u_{A_{\parallel}}-\beta)^2}\Big[1-\frac{u_{A_{\perp}}^2}{\gamma^2(1-\beta u_{A_{\parallel}})^2}\Big]+k u_{f} \frac{1-\beta u_{f}}{(1-\beta u_{A_{\parallel}})^2}\bigg]\Bigg] \Bigg\}, \nonumber
\end{align}
and 

\begin{align}
\Pi(k,\omega)\equiv-\frac{\beta}{\gamma}B_{z}^{(0)}\frac{\omega}{(1-\beta u_{A_{\parallel}})^2}\frac{1-u_{A}^2}{(u_{A_{\parallel}}-\beta)^2}&\frac{(u_{f}-u_{A_{\parallel}})\big[(u_{f}+u_{A_{\parallel}})(1+\beta^2)-2\beta(1+u_{f}u_{A_{\parallel}})\big]}{(1-\beta u_{f})^2(1-\beta u_{s})^2} 
 \\
&\bigg[\frac{u_{s}(u_{s}-\beta)(1-\beta u_{s})}{\omega^2-k^2 u_{s}^2}-\frac{u_{f}(u_{f}-\beta)(1-\beta u_{f})}{\omega^2-k^2 u_{f}^2}\bigg]. \nonumber
\end{align}
\end{widetext} \


\end{document}